\documentclass[%
reprint,
superscriptaddress,
amsmath,
amssymb,
aps,
twocolumn,
pra,
floatfix
]{revtex4-2}

\usepackage{xcolor}
\usepackage[colorlinks,linkcolor=blue,citecolor=blue,urlcolor=blue]{hyperref} 
\usepackage{graphicx}
\usepackage[english]{babel}
\usepackage[utf8]{inputenc}
\usepackage{amssymb}
\usepackage{amsmath}
\usepackage[capitalize]{cleveref}
\usepackage{braket}
\usepackage{siunitx}
\usepackage{color}
\usepackage{csquotes}
\usepackage{newtxtext}
\usepackage{newtxmath}
\usepackage{soul}
\usepackage{shellesc}
\usepackage{nicematrix}

\makeatletter 
\renewcommand{\fnum@figure}{FIG~\thefigure}
\makeatother

\newcommand{\imunit}{\mathrm i}
\newcommand{\diffd}{\mathrm d}

\newcommand{\affHAN}{\address{Institut f{\"u}r Quantenoptik, Leibniz Universit{\"a}t Hannover, Welfengarten 1, D-30167 Hannover, Germany}}
\newcommand{\affULM}{\address{Institut f{\"u}r Quantenphysik and Center for Integrated Quantum Science and Technology (IQ\textsuperscript{ST}), Universit{\"a}t Ulm, Albert-Einstein-Allee 11, D-89069 Ulm, Germany}}
\newcommand{\affTUDa}{\address{Technische Universit{\"a}t Darmstadt, Fachbereich Physik, Institut f{\"u}r Angewandte Physik, Schlossgartenstr. 7, D-64289 Darmstadt, Germany}}
\newcommand{\affDLR}{\address{Institute of Quantum Technologies, German Aerospace Center (DLR), Wilhelm-Runge-Stra{\ss}e 10, D-89081 Ulm, Germany}}

\begin{document}
\title{Bragg-diffraction-induced imperfections of the signal in retroreflective atom interferometers}

\author{Jens Jenewein}%
\email{jens.jenewein@uni-ulm.de;info@jensjenewein.de}
\affULM
\author{Sabrina Hartmann}
\affULM
\author{Albert Roura}
\affDLR
\author{Enno Giese}
\affTUDa
\affHAN


\collaboration{This article has been published in \href{https://doi.org/10.1103/PhysRevA.105.063316}{Physical Review A \textbf{105}, 063316 [2022]}}
\begin{abstract}
We present a detailed study of the effects of imperfect atom-optical manipulation in Bragg-based light-pulse atom interferometers. 
Off-resonant higher-order diffraction leads to population loss, spurious interferometer paths, and diffraction phases.
In a path-dependent formalism, we study numerically various effects and analyze the interference signal caused by an external phase or gravity.
We compare first-order single and double Bragg diffraction in retroreflective setups.
In double Bragg diffraction, phase imperfections lead to a beating due to three-path interference.
Some effects of diffraction phases can be avoided by adding the population of the outer exit ports of double diffraction.
\end{abstract}
\maketitle

\section{\label{sec:Introduction}Introduction}

Light-pulse atom interferometry \cite{kasevich_atomic_1991,Kleinert2015} has demonstrated a great potential for precision measurements in both fundamental physics and more practical applications \cite{Bongs2019}. In this context, atomic Bragg diffraction has become a versatile tool for enhancing the sensitivity of atom interferometers through techniques like sequential pulses \cite{PhysRevLett.103.080405,PhysRevLett.114.063002,kovachy2015quantum} as well as double \cite{Ahlers2016,kuber_experimental_2016,giese2013double} and higher-order diffraction \cite{muller_atom_2008,siems_analytic_2020,chiow_102hbark_2011,gebbe_twin-lattice_2021}.
Because its efficiency crucially depends on the pulse duration, Bragg diffraction favors an intermediate regime where velocity selectivity does not dominate, but spurious diffraction orders may occur \cite{Szigeti_2012,Hartmann2020a,neumann_aberrations_2021,manna_nonadiabatic_2020}.
In this article, we study these effects in single and double Bragg interferometry.
We examine not only deleterious interferometer paths \cite{lu_competition_2018,altin_precision_2013,plotkin-swing_three-path_2018,parker_controlling_2016,lu_competition_2018,he_phase_2021}, but also phase errors induced by imperfect atom-optical elements.

Due to the finite momentum spread of the atomic wave function, there is a varying Doppler detuning across the momentum distribution, which implies a loss of diffraction efficiency.
This behavior is known as velocity selectivity \cite{Szigeti_2012,giltner_theoretical_1995,durr_acceptance_1999} and its effects are suppressed for short pulse durations.
In contrast to Raman diffraction \cite{kasevich_atomic_1991,moler_theoretical_1992}, however, (single) Bragg diffraction \cite{torii_mach-zehnder_2000,giese2015mechanisms} from an optical lattice is not a perfect two-level system, especially in the so-called Raman-Nath (or Kapitza-Dirac) regime \cite{muller_atom-wave_2008,gould1986diffraction}.
Therefore, diffraction into off-resonant higher-order momenta can become relevant \cite{neumann_aberrations_2021,manna_nonadiabatic_2020,parker_controlling_2016,he_phase_2021}.
Since velocity selectivity and spurious higher-order diffraction are two competing effects, an intermediate regime is advised \cite{beguin_characterization_2021,muller_atom-wave_2008}. 
For a careful assessment of such a regime, we not only focus on resonant first-order single Bragg diffraction, but also study double Bragg diffraction from two optical lattices propagating in opposite directions.
Both atom-optical diffraction mechanisms have three major effects of imperfections: (i) The loss of population on desired interferometer paths, (ii) the emergence of spurious paths and (iii) additional phase errors.

Phases observed in an atom-interferometric signal that arise from light-matter interaction and differ from those that would be induced by ideal laser pulses are called diffraction phases.
Any detuning introduces such phases, e.g., ac Stark shifts in Raman diffraction~\cite{moler_theoretical_1992} or couplings to additional and off-resonant light fields causing the two-photon light shift \cite{carraz_phase_2012,gauguet_off-resonant_2008,giese_light_2016}.
Resonant higher-order diffraction in large-momentum transfer techniques also gives rise to additional phases \cite{muller_atom-wave_2008,siems_analytic_2020,gochnauer_bloch-band_2019,estey_high-resolution_2015}, since such a configuration constitutes only an approximate two-level system and intermediate momentum states are relevant.
Additional contributions to the interferometer phase arise when atoms drop during the pulses.
These phase shifts depend on the pulse duration and gravity \cite{li_raman_2015,bertoldi_phase_2019,antoine_matter_2006,peters_high-precision_2001} and can be calculated perturbatively in a two-level system.
In this context, they can be interpreted as pulse imperfections and diffraction phases as well.
Because even resonant first-order Bragg diffraction is inherently not a two-level system, population can be lost to higher off-resonant diffraction orders \cite{Hartmann2020a,neumann_aberrations_2021,beguin_characterization_2021,parker_controlling_2016}.
Such processes are of particular relevance for double Bragg diffraction, where a central path arises with significant population \cite{fitzek_universal_2020}.
Consequently, three-path interference \cite{plotkin-swing_three-path_2018,gupta_contrast_2002} becomes important for double Bragg diffraction, but is also observed in double Raman diffraction \cite{he_phase_2021}.
While losses to higher diffraction orders in double Bragg diffraction were at the focus of Ref.~\cite{Hartmann2020a}, the study focused on the diffraction efficiency.
Only contrast and amplitude of two-path interference was studied and the effects of diffraction phases were not discussed in detail.

We extend these studies by combining multiple sources of errors to obtain a more complete picture.
For first-order single Bragg diffraction, we show that there is indeed loss of atoms, but effectively no spurious paths contribute to the interference signal of an Mach-Zehnder interferometer in an intermediate regime.
Due to the symmetry of the atom-optical interactions in such an interferometer, all phase imperfections cancel out.
Including a gravitational acceleration during the pulse, this symmetry is broken and we observe phases that arise from the finite pulse durations and acceleration.
Similar to the Raman-Nath regime \cite{plotkin-swing_three-path_2018,he_measuring_2022}, in double Bragg diffraction there is a significant contribution of a third path to the interference and a beating pattern arises.
We observe an asymmetric phase shift between the two outer exit ports (which has been experimentally observed in Refs.~\cite{Ahlers2016,he_phase_2021}) and show that it cancels if both populations are added.
In first-order double diffraction, effects of gravity cannot be trivially compensated by chirping, since in retroreflection both lattices are accelerated in opposite directions.
One possible way to overcome these issues is to add a third laser frequency \cite{Malossi_Double_2010,he_phase_2021}, but at the cost of laser phases contributing to the interference signal.
In contrast, double diffraction in a retroreflective geometry is independent of laser phase noise to lowest order.
However, it is limited to microgravity or quasi-horizontal configurations \cite{zhou_test_2015,Ahlers2016}.
We observe minimal gravitational effects in an experiment with the mirror at the apex.

Our numerical simulations are built on a software suite based on a path-dependent description of light-pulse atom interferometry.
It can be used to model atom interferometers in linear gravity including finite pulse durations and arbitrary pulse envelopes, velocity selectivity, as well as higher-order diffraction.

The structure of the article is as follows:
We introduce our path-based approach to the description of light-pulse atom interferometers in Sec.~\ref{sec:model}.
Excluding gravity from the discussion, we investigate in Sec.~\ref{sec:nograv} the population of the individual paths of Mach-Zehnder interferometers that contribute to the interference signal as well as the overlap of the wave functions that propagated along the different paths.
This way, we are able to examine phase contributions that arise from imperfect beam splitters and mirrors, study the interference signal and analyze the beating pattern that arises in double Bragg diffraction.
In Sec.~\ref{sec:gravity}, we extend our results to nonvanishing gravitational acceleration and observe the effects of phase errors on the interferometer signal.
We conclude in Sec.~\ref{sec:conclusion} and summarize in Appendix~\ref{app:diffODE} the numerical approach involved in describing the diffraction process.

\section{Model}
\label{sec:model}

Light-pulse atom interferometers consist of a sequence of light pulses inducing diffraction from optical gratings, that sandwich the center-of-mass evolution in an external potential.
We describe the latter by the operator $\hat U(t) = \exp \left( -\text{i} \hat H_0 t /\hbar \right)$, where the Hamiltonian $\hat{H}_0 = \hat{p}^2/(2M) + M  a \hat{z}$ includes the projection $a$ of the gravitational acceleration on the direction of the diffracting light beams.
Here, $M$ denotes the atomic mass,  as well as $\hat p$ and $\hat z$ its momentum and position, which obey the commutation relation $[\hat z, \hat p] = \imunit \hbar$.
The $j$th diffraction process is described by the operator $\hat{G}^{(j)}$ in an interaction picture with respect to $\hat H_0$, initiated at the beginning of the pulse.
Throughout this article, we assume Gaussian light pulses of width $\Delta \tau$  that induce diffraction of the atomic wave packets and solve numerically the Schrödinger equation describing different diffraction mechanisms. 
Details are given in Appendix~\ref{app:diffODE}.

The operator sequence  $\hat{U}_{\mathrm{AI}}$ describing an atom interferometer with three pulses $j = 1,2,3$ reads in the Schrödinger picture
\begin{equation}
\label{eq:U_AI}
\begin{split}
  \hat{U}_{\mathrm{AI}}=  &\hat{U}(t_\text{d}) \hat{U}^\dagger(t_3)\hat{G}^{(3)}\hat{U}(t_\text{3}) \\ &\hat{U}^\dagger(t_2)\hat{G}^{(2)}\hat{U}(t_\text{2})
    \hat{U}^\dagger(t_1)\hat{G}^{(1)}\hat{U}(t_\text{1}),
    \end{split}
\end{equation}
where $t_\text{d}$ describes the time of detection and $t_j$ the beginning of each pulse.
All unitary transformations between different pictures have been absorbed into the evolution operators $\hat{U}(t_j)$.
The operator $\hat U(t_\text{d})$ leads to a global phase in momentum representation and a momentum shift of the output by $- M a t_\text{d}$.
Since the detection is performed by measuring all atoms in a certain exit port defined by a momentum interval, the phase is irrelevant and the shift can be absorbed in the definition of the exit port.
Hence, we can ignore the action of $\hat U(t_\text{d})$ in the following, but are restricted to momentum representation.

\begin{figure}[t]
\includegraphics[width=\columnwidth]{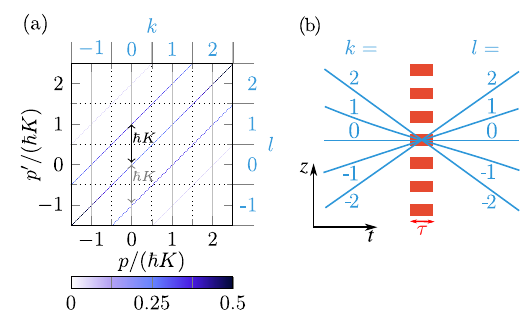}
\caption{
(a) Transition function $|G(p^\prime,p)|^2$ of a SBD beam splitter with pulse duration $\Delta \tau = \SI{20}{\micro \second}$, connecting the initial momentum $p$ to the final momentum $p^\prime$.
Only transitions with a momentum difference that is an integer multiple of $\hbar K$ occur, as indicated by the antidiagonals with nonvanishing diffraction probability.
It is therefore possible to label the central momenta by integer numbers $k$ and $l$ (blue axis labels), where $|l-k|$ is proportional to the momentum transfer and denotes the diffraction order.
(b) Schematic of the diffraction process:
Different incoming momenta denoted by $k$ correspond to different slopes in a space-time diagram and impinge on an optical grating of duration $\tau$ (red). They are diffracted into outgoing momenta $l$ that become spatially separated over the course of time.
}
\label{fig:transitionFunction}
\end{figure}
\begin{figure*}
	\centering
	\includegraphics[width=1\textwidth]{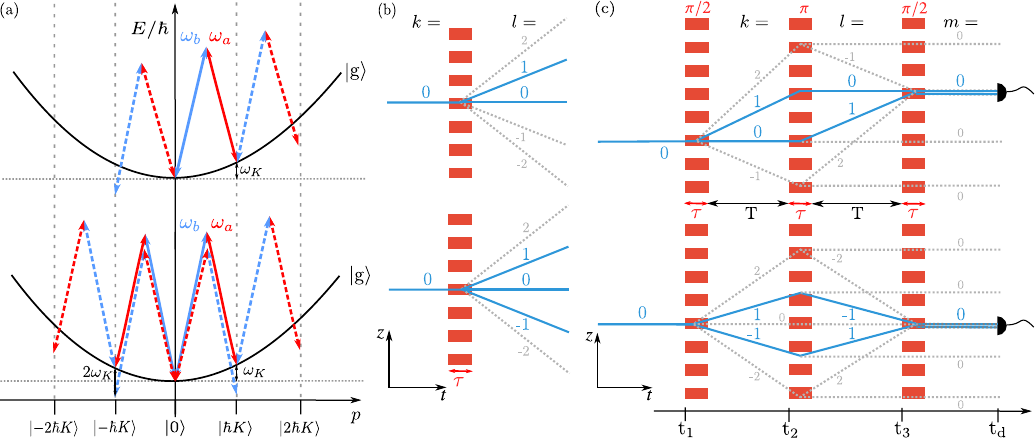}
	\caption{
	Diffraction mechanisms and interferometer geometries of SBD (top row) and DBD (bottom row).
	(a) Dispersion relations associated with the diffraction process.
	If the frequency difference $\omega_b - \omega_a = \omega_K$ between the blue and red counterpopagating lasers holds, an atom at rest experiences resonant first-order diffraction as shown by the blue and red solid lines.
	The dashed lines indicate off-resonant processes for SBD that correspond to higher diffraction orders.
	For DBD, two processes occur simultaneously in opposite directions.
	In this case, the momenta associated with first order are now coupled resonantly and off-resonantly.
	(b) Space-time diagrams for an incoming momentum $k=0$.
	The blue lines show the resonant momenta, the dotted gray lines off-resonant higher-order diffraction.
	For SBD we expect two outgoing resonant paths with $l = 0$ and $l = 1$.
	For DBD there are in principle three outgoing resonant paths with  $l = 1$, $l = 0$ and $l = -1$.
	However, for an ideal beam splitter, the central branch is not populated.
	(c) Mach-Zehnder interferometer sequences for a vanishing incoming momentum with beam-splitter pulses at times $t_1$ and $t_3$ and a mirror pulse at time $t_2$.
	The pulses have a duration of $\tau$ and are separated by a time $T$.
	Each path through the interferometer is uniquely specified by a triple of diffraction indices $(m,l,k)$, connected to different momenta.
	The desired paths , which are generated by resonant first-order diffraction, are drawn with blue lines and spurious ones with dashed lines.
	Each exit port is chosen via a spatially resolved measurement.
	}
	\label{fig:interferometer}
\end{figure*}

Calculating the output wave function $\psi_\text{out}(p')$ after diffraction from an input $\psi_\text{in}(p)$, we introduce in analogy to Ref.~\cite{Hartmann2020a} the transition function in momentum representation $G^{(j)}(p',p) \equiv \braket{p' | \hat{G}^{(j)} | p}$ via
\begin{equation}
	\psi_\text{out}(p') = \int \, G^{(j)}(p',p) \, \psi_\text{in}(p)  \text{d} p.
	\label{eq:psi_final}
\end{equation}
In Fig.~\ref{fig:transitionFunction}\,(a) we display as an example the probability for a transition from input momentum $p$ to output momentum $p'$ obtained from a numerical solution of single Bragg diffraction described by the Schr\"odinger equation in Appendix~\ref{app:diffODE}.
The clear structure of antidiagonals that are separated by the recoil momentum $\hbar K$ shows distinct diffraction orders.
We observe resonant processes as indicated by the black arrow highlighting $\ket{0} \rightarrow \ket{\hbar K}$, but also off-resonant ones $\ket{0} \rightarrow \ket{-\hbar K}$ indicated by the gray arrow.
The physical reason for this discrete structure is explained in the context of Fig.~\ref{fig:interferometer}.

When we calculate the momentum representation $\braket{p' | \hat{U}^\dagger(t_j)\hat{G}^{(j)}\hat{U}(t_j) | p}$ of the sandwiched transition operators that arise in Eq.~\eqref{eq:U_AI}, we find
\begin{equation}
    \exp\left(  \imunit \frac{p-p'}{2 \hbar} a t_j^2  + \imunit \frac{p^{\prime 2} - p^2}{2M\hbar } t_j \right)   G^{(j)}(p' - M a t_j,p - M a t_j).
    \label{eq:diffOpMom}
\end{equation}
Hence, the dropping of the atoms prior to diffraction is taken into account by the argument $p-M a t_j$.

Since input and output momenta are separated by multiples of $\hbar K$, we introduce $p' = p_0 + l \hbar K$ and $p = p_0 + k \hbar K$, where $k,l \in \mathbf{Z}$.
The integers $l,k$  are also shown in Fig.~\ref{fig:transitionFunction}\,(a) on the top and right in blue.
The difference $|l - k|$ defines the diffraction order.
In general, the diffraction process produces a superposition of momentum states which, in turn, leads to a separation of atomic trajectories in position space upon time evolution.

Using the labels $k,l$, we rewrite Eq.~\eqref{eq:diffOpMom} as
\begin{equation}
 \operatorname{e}^{\imunit \varphi_{l,k}^{(j)}}   G^{(j)}(p_0 +l \hbar K- M a t_j,p_0+ k \hbar K-M a t_j)
\end{equation}
with phase
\begin{equation}
     \varphi_{l,k}^{(j)} (p_0)=  \left(l^2 - k^2\right)\omega_K t_j  + (l-k) \omega_\mathrm{D}(p_0) t_j+ (k-l)  \frac 1 2 K a t_j^2.
\end{equation}
Here, we introduced the \textit{Doppler detuning} $\omega_\mathrm{D}(p_0)= p_0 K / M$ and the \textit{recoil frequency} $\omega_K = \hbar K^2 /(2M)$, respectively. 

We now calculate the time evolution of the whole interferometer sequence $\bra{p_0+ m \hbar K }\hat{U}_{\mathrm{AI}}\ket{p_0} = U_{m}(p_0)$ in momentum representation.
Note that the index $m$ denotes the \emph{exit port}.
We arrive at the expression
\begin{equation}
\begin{split}
    U_{m}(p_0) = \sum\limits_{k,l}& \operatorname{e}^{ \imunit \varphi_{m,l,k} } G^{(3)}_{m,l}(p_0 -M a t_3) \\
    & \times G^{(2)}_{l,k}(p_0 -M a t_2) G^{(1)}_{k,0}(p_0 -M a t_1).
    \end{split}
\end{equation}
The sum over $k$ and $l$ gives rise to a superposition of different paths that end in the same exit port $m$.
We have used the abbreviation $G^{(j)}_{l,k}(p_0 - M a t_j) \equiv G^{(j)}(p_0 +l \hbar K- M a t_j,p_0+ k \hbar K-M a t_j)$ to highlight the labels of the initial and final momentum and introduced the phase
\begin{equation}
    \varphi_{m,l,k}=\varphi_{m,l}^{(3)} +\varphi_{l,k}^{(2)} +\varphi_{k,0}^{(1)}
    \label{eq:definitionIntPhase}
\end{equation}
that is associated with a path.

We assign a wave function $\psi_{m,l,k}(p_0+m \hbar K)$ to each path (dropping $m \hbar K$ in the argument in the following), which is identified by the tuple $(m,l,k) \in \mathbb Z^3$ and has the form
\begin{equation}
\begin{split}
    \psi_{m,l,k}(p_0) = \operatorname{e}^{\imunit \varphi_{m,l,k}} &G^{(3)}_{m,l}(p_0 -M a t_3)G^{(2)}_{l,k}(p_0 - M a t_2) \\
    &  \times G^{(1)}_{k,0}(p_0 -M a t_1) \psi_\text{in}(p_0),
    \end{split}
    \label{eq:defPsi}
\end{equation}
where $\psi_{\mathrm{in}}(p_0)$ denotes the initial wave function in momentum representation.
Throughout this article we assume for the initial wave function a Gaussian distribution
\begin{equation}
\psi_{\mathrm{in}}(p_0) \sim \exp[-(p_0 - \bar{p}_0)^2 / (4 \Delta \wp^2)]
\label{eq:defIniWF}
\end{equation}
of width $\Delta \wp \ll \hbar K$ around the central momentum $\bar p_0$. 
The function $\psi_{m,l,k}(p_0)$ is associated with the subsequent diffraction processes leading to accumulated momentum transfers corresponding to $k,l$ and $m$, and describes the propagation along a unique path through the interferometer.
In fact, without gravity and for an initially vanishing central momentum, $\bar p_0 = 0$, the accumulated momentum transfer at each stage coincides with the central momentum of the wave function at that time.
 
In light-pulse atom interferometers, the momentum transfer is typically induced by diffraction from optical light fields. 
For \textit{single Bragg diffraction} (SBD), one optical grating consisting of two counterpropagating lasers causes diffraction as shown in the top of Fig.~\ref{fig:interferometer}\,(a):
For instance, the atom absorbs light from a blue laser beam of frequency $\omega_b = c k_b$ and experiences an associated momentum kick $\hbar k_b$. 
The counterpropagating (red) laser beam leads to stimulated emission of frequency $\omega_a= c k_a$ and a decay back to the ground state.
A momentum kick $\hbar k_a$ in the same direction is associated with this transition. 
Hence, the overall momentum $\hbar K \equiv \hbar k_b + \hbar k_a$ is transferred. 
The frequency difference between the lasers $\omega_b - \omega_a = \Delta \omega$ determines the resonant diffraction order based on energy and momentum conservation. 
Here, we focus on resonant first-order diffraction.
Depicted is a resonant transition between the momentum states $\Ket{0}$ and $\Ket{\hbar K}$.
The corresponding resonance condition $\Delta \omega = \omega_K$ for such a first-order process shows that the absorbed energy corresponds exactly to the gained kinetic energy.
The resonant momenta are connected by solid arrows, whereas the off-resonant higher-order diffraction is drawn as dashed arrows.
This effective two-photon process is discussed in more detail in Appendix~\ref{app:diffODE} that also contains the corresponding set of differential equations.

The same principle applies to \textit{double Bragg diffraction} (DBD) induced by adding a second optical grating, for which the direction of the two lasers is reversed~\cite{giese2013double,kuber_experimental_2016,Ahlers2016}.
Both gratings lead to diffraction into opposite directions, as shown in the lower part of Fig.~\ref{fig:interferometer}\,(a). 
Each pair induces a resonant coupling of two momenta in one direction, whereas the opposite momentum is coupled off-resonantly.
The respective differential equations for DBD are also provided in Appendix~\ref{app:diffODE}.

The resonance condition directly defines the preferred incoming and outgoing momenta, for which we use the labels $k$ and $l$.
The choice $\Delta \omega = \omega_K$ depicted in the top of Fig.~\ref{fig:interferometer}\,(a) for SBD corresponds for an incoming momentum $k=0$ to diffraction into the orders $l = 0,1$, as shown in the top of Fig.~\ref{fig:interferometer}\,(b) by blue lines.
Dashed gray lines denote off-resonant higher diffraction orders.
The same resonance condition in DBD gives rise to resonant moment; the incoming momentum $k = 0$ is resonantly connected to $l = \pm 1,0$, depicted in the lower part of Fig.~\ref{fig:interferometer}\,(b). 
Off-resonant higher-order processes such as $l = \pm 2$ are suppressed.

If off-resonant orders can be neglected, SBD can be interpreted as an effective Rabi oscillation between two resonant momenta. 
The duration, i.e., the pulse area of the interaction determines whether an equal superposition of momenta is generated, or the population of the momentum states is inverted.
The former is referred to as \textit{beam splitter} or $\pi/2$ pulse, the latter as \textit{mirror} or $\pi$ pulse.
In contrast, the effective Rabi oscillations in DBD are those of a three-level system.
A beam splitter corresponds to the generation of a superposition of momenta $\pm 1$.
A mirror corresponds to the transition from $k = \pm 1$ to $l = \mp 1$.
The definition of the pulse area is provided in Appendix~\ref{app:diffODE}.

With these atom-optical elements we build a Mach-Zehnder atom interferometer, which consists of a $\pi/2$ pulse that generates a superposition of two paths, a redirecting mirror pulse, and a final $\pi/2$ pulse that recombines the to paths to observe interference.
Figure~\ref{fig:interferometer}\,(c) shows such an interferometer for both SBD (top) and DBD (bottom).
The resonant paths are drawn by blue lines.
The most relevant off-resonant diffraction orders are denoted by dashed gray lines.

The figure shows many off-resonant paths exiting the interferometer with $m=0$.
To prevent the detection of most of those paths, we place a detector at a certain position after the detection time $t_{\mathrm d}$, as highlighted by the detector symbol.
This combination of momentum-dependent and position-dependent detection at intermediate times leads to a cleaner signal.
However, since the paths that end in the detector can be associated with a tuple $(m,l,k)$, we can still use the formalism developed above, where we ignored $\hat U(t_\text{d})$.
The final wave function is a superposition of all the wave functions $\psi_{m,l,k}$ that end in the detector in the figure.
Since no state labeling is possible for Bragg, in contrast to Raman~\cite{lu_competition_2018}, we assume a spatially resolved measurement of the exit ports.
Thus, all interfering wave functions are defined by $m$ as well as $l+k$ , which respectively characterize the central momentum and the central position after the last beam splitter. The interferometer times $T$ that we are primarily interested in will typically guarantee that the spatial separation between wave functions with different values of $l + k$, which is proportional to $(\hbar K / M) T$, is larger than the size of the atomic cloud. Moreover, the possible overlap of one of the main ports and a spurious port with a momentum differing by a multiple of $\hbar K$ can be avoided by adjusting the time until detection, but would in any case give no coherent contribution to the interferometric signal.

For SBD, the upper resonant path reads $(0,0,1)$ and the lower resonant path $(0,1,0)$ which are indicated by blue lines in the top of Fig.~\ref{fig:interferometer}\,(c).
For example, the gray dotted off-resonant paths $ (0,-1,2)$ and $ (0,2,-1)$ are also detected (since $m=0$ and $l+k=1$) and a result of the imperfect nature of higher-order diffraction.
We now introduce the times $t_j=(j - 1)(T+\tau)$ for $j = 1,2,3$, where $\tau$ is the overall duration of the diffracting pulses and $T$ the interrogation time of the interferometer.
We find with the help of Eq.~\eqref{eq:definitionIntPhase} the phase difference  
\begin{align}
    \varphi_{0,0,1} - \varphi_{0,1,0} = -K a (T + \tau)^2.
    \label{eq:kgtsquaredsingle}
\end{align}
associated with the two resonant paths.
It can be connected to the usual phase measured by gravimeters.
However, note that the transition functions $G^{(j)}_{l,k}$ may introduce additional phases to $ \psi_{m,l,k}$ that might depend on the momentum.
We will analyze their effect in the sections below.

In DBD, the two resonant paths are $ (0,-1,1)$ and $ (0,1,-1)$, as shown in the bottom of Fig.~\ref{fig:interferometer}\,(c).
The phase difference associated with these paths is
\begin{align}
    \varphi_{0,-1,1} - \varphi_{0,1,-1} =  - 2K a (T + \tau)^2,
    \label{eq:kgtsquareddouble}
\end{align}
but further contributions may arise due to the phases of the transition functions $G^{(j)}_{l,k}$.
This phase difference captures one feature of DBD: the enhancement of sensitivity caused by the factor of two compared with SBD.
Off-resonant paths $m=0$ and $m+l+k=0$ are also detected, for example $(0,-2,2)$ and $ (0,2,-2)$.
Moreover, the detected center path $(0,0,0)$ plays a special role.
Even though in an ideal and resonant three-level system it is not populated, it still has to be considered a quasiresonant path.

\section{Effects for vanishing gravity}
\label{sec:nograv}

To simplify the discussion, we first study the interference signal generated by resonant first-order diffraction for vanishing gravity, that is $a=0$.
Already in this case, we observe some key features of imperfect diffraction and main differences between SBD and DBD.
We solve the transition functions in momentum representation based on the differential equations provided in Appendix~\ref{app:diffODE} for both SBD and DBD.
For that, we assume Gaussian pulses of variable temporal width $\Delta \tau$ that is the same for all three diffracting pulses.
For all simulations, we use initial momentum widths $\Delta \wp$ that are comparatively large for experiments with Bose-Einstein condensates and are detailed in the respective figures.
However, it is wide enough to observe significant effects from velocity selectivity.

\subsection{Exit port contribution of individual paths}
\label{subsec:contributions}

To characterize the influence of the individual paths through the interferometer, we define the contribution $P_{m,l,k}$ of path $(m,l,k)$ to port $m$ by
\begin{equation}
	P_{m,l,k} = \int |\psi_{m,l,k}(p)|^2 \diffd p .
	\label{eq:contribution}
\end{equation}
The limits of integration are chosen $\pm \hbar K / 2$ around the center of the wave packet $\psi_{m,l,k}$ from Eq.~\eqref{eq:defPsi}.

\begin{figure}[h]
\includegraphics[width = \columnwidth]{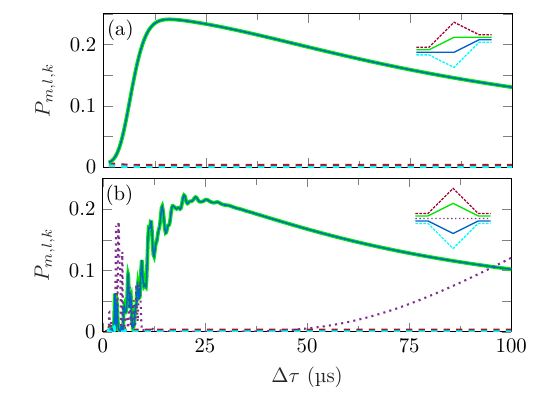}
\caption{
    Population $P_{m,l,k}$ for each individual path that ends in exit port $m$.
    The legends on the top right show the respective color-coded path and can be identified with the indices shown in Fig.~\ref{fig:interferometer}\,(c).
    The desired paths are drawn with continuous lines, spurious paths with dashed lines.
    Panel (a) shows that for SBD only the two resonant paths contribute to the interference signal and they have the same population.
    For short times (in the Raman-Nath regime), population is lost to spurious paths which, however, do not end in the exit port.
    For long times (in the deep Bragg regime), population is lost due to velocity selectivity.
    Panel (b) shows that the situation differs for DBD.
    While both desired branches are symmetrically populated and again show a decreasing population in the Bragg regime, the behavior is richer in the Raman-Nath regime.
    Moreover, velocity selectivity leads to a significant population of the central path  (dotted line) that ends in the central exit port.
    For these simulations we use an initial Gaussian momentum width of $\Delta \wp = 0.05 \hbar K$ and an interferometer time of $T = \SI{0.1}{\second}$.
}
\label{fig:Contributions}
\end{figure}
For the case of SBD, we observe in Fig.~\ref{fig:Contributions}\,(a) a decreasing population for increasing durations $\Delta \tau$, i.e., in the Bragg regime. The parameters for our simulations can be seen in Appendix~\ref{app:diffODE}.
The reason for this decrease is velocity selectivity: the atoms on the wings of the momentum distribution are Doppler detuned from resonance, an effect more prominent for decreasing Rabi frequencies.
On the other hand, small times  correspond to the Raman-Nath regime, where the diffraction is governed by loss to higher orders. 
Moreover, we observe $P_{0,1,0} = P_{0,0,1}$ so that both paths contribute equally to the signal. 
Population of higher-order paths $P_{0,-1,2}$ and $P_{0,2,-1}$ as indicated by the dotted lines in Fig.~\ref{fig:interferometer}\,(a) is completely negligible.
The reason is obvious: to end in the exit port, more than one subsequent higher-order diffraction process is necessary, which drastically reduces its probability.
Therefore, only the two resonant paths lead to a relevant contribution, if the interrogation time $T$ as well as the detection time $t_\text{d}$ are sufficiently large to separate different exit ports.

For DBD, we observe in Fig.~\ref{fig:Contributions}\,(b) qualitatively the same behavior for small and large times.
However, in the quasi-Bragg regime at intermediate durations where the off-resonant coupling of resonant states becomes increasingly important, a richer structure emerges.
Nevertheless, the contributions of the resonant paths to port $m=0$ are still the same, i.e., $P_{0,-1,1} = P_{0,1,-1}$.
For the outer exit ports $m = \pm 1$ (not shown in the figure), the contribution of the upper path to the lower port is equal to the contribution of the lower path to the upper port $P_{1,-1,1} = P_{-1,1,-1}$.
Similarly, we find $P_{-1,-1,1} = P_{1,1,-1}$.
This effect is a direct consequence of velocity selectivity:
For example, the upper path ending in the upper exit port has three changes of momentum, whereas the lower path only has two changes in momentum, each one associated with a velocity selection.

The contributions $P_{0,-2,2}$ and $P_{0,2,-2}$ of spurious paths to the central exit port are negligible for the same reason as in SBD: two higher-order processes are necessary, as indicated in Fig.~\ref{fig:interferometer}\,(c) by dotted lines.
However, in contrast to SBD, the contribution $P_{0,0,0}$ of the central path is significant and must be included.
This path is not off-resonant, but only depopulated, since it is part of the three-level system.
However, for long velocity-selective pulses, an increasingly important population remains on the central path, as shown in Fig.~\ref{fig:Contributions}\,(b) by the dotted line.
As a consequence, there are three paths contributing to the interference signal.

\subsection{Overlap of individual paths}
\label{subsec:overlap}

\begin{figure}[h]
\includegraphics[width = \columnwidth]{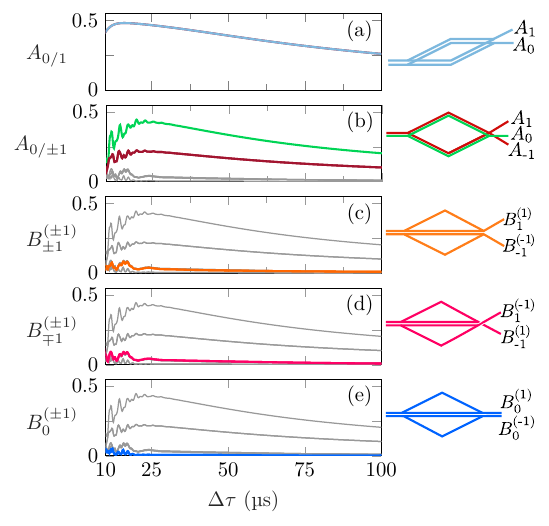}
\caption{
    Moduli $A_m$ and $B_m^{(\pm 1)}$ of the overlap in each exit port $m$ of the wave functions that propagated along different paths.
    While $A_m$ corresponds to the overlap of the two desired paths, $B_m^{(\pm 1)}$ is the overlap between one spurious and one desired path.
    The legend on the right connects each overlap to those two paths and to an exit port through a color code.
    The gray lines give the overlap of other panels for reference.
    Panel (a) shows that the overlap of the two paths in both exit ports is the same for SBD.
    Similar to the populations from Fig.~\ref{fig:Contributions}, we observe effects of the Raman-Nath regime and velocity selectivity.
    Panel (b) shows an analog behavior for the overlap of the two desired paths in DBD.
    Note that the modulus of the overlaps in the outer exit ports is the same, i.e., $A_{-1}=A_1$.
    However, in DBD we also observe a significant overlap between the central and the outer paths in the outer exit ports; see panels (c) and (d).
    Such spurious overlaps are strongly suppressed in the central exit port, see panel (e).
    For these simulations we use an initial Gaussian momentum width of $\Delta \wp = 0.05 \hbar K$ and an interferometer time of $T = \SI{0.1}{\second}$.
    }
\label{fig:overlapAbs}
\end{figure}

The interference signal in the exit ports is not only determined by their contribution, but also by the overlap of the individual wave functions ending in a particular exit port.
In SBD, we saw that only two paths yield significant population.
To observe a phase variation in the signal, we artificially imprint the \textit{external phase} $\pm \vartheta_{\text{ext}}/2$ onto each one of them.
Furthermore, we assume no gravity, so Eq.~\eqref{eq:kgtsquaredsingle} (and Eq.~\eqref{eq:kgtsquareddouble}) vanish.
Hence, the wave function in momentum representation takes the form
\begin{equation}
   \Psi_m(p) = \mathrm{e}^{\imunit \vartheta_{\mathrm{ext}} / 2} \psi_{m,0,1} + \mathrm{e}^{-\imunit \vartheta_{\mathrm{ext}} / 2} \psi_{m,1,0} .
\end{equation}
Thus, we write the interference signal $I_m = \int  |\Psi_m(p)|^2 \diffd p$ in exit port $m$ as
\begin{equation}
    I_m = P_{m,1,0} + P_{m,0,1} + A_m \cos(\vartheta_{\mathrm{ext}}- \alpha_m)
    \label{eq:defSignalSBD}
\end{equation}
with the populations from Eq.~\eqref{eq:contribution} discussed above.
Here, we define the overlap between two paths
\begin{equation}
     \int \psi_{m,0,1}^\ast \psi_{m,1,0} \diffd p = \frac 1 2 A_m \exp{(\imunit \alpha_m)}
    \label{eq:defOverlapSBD}
\end{equation}
with amplitude $A_m$ that leads to contrast $A_m / (P_{m,1,0} + P_{m,0,1})$.
The phase of the overlap $\alpha_m$ will be discussed in the section below.
As we see in Fig.~\ref{fig:overlapAbs}\,(a), the amplitude of the overlap decreases due to velocity selectivity and is the same in both exit ports, i.e., $A_{0} = A_1$.
When we recall from Fig.~\ref{fig:Contributions}\,(a) in addition that $P_{0,1,0} = P_{0,0,1}$, it is evident that the contrast is the same in both exit ports.
Because velocity selectivity acts both on $A_m$ and $P_{m,1/0,0/1}$, the contrast for port 0 is unity throughout all regimes.

As shown above, in DBD three paths contribute significantly to the exit ports.
We again introduce a varying interference signal by adding a phase factor to the outer paths, but none to the central path.
Hence, a superposition
\begin{equation}
   \Psi_m(p) = \mathrm{e}^{\imunit \vartheta_{\mathrm{ext}} / 2} \psi_{m,-1,1} + \psi_{m,0,0}+ \mathrm{e}^{-\imunit \vartheta_{\mathrm{ext}} / 2} \psi_{m,1,-1} 
   \label{eq:defSignalDBDTwoPath}
\end{equation}
is detected and leads to the signal
\begin{equation}
\begin{split}
    I_m=&\, P_{m,-1,1} + P_{m,0,0} + P_{m,1,-1} + A_m \cos \left( \vartheta_{\mathrm{ext}} - \alpha_m \right) \\
    & + B_m^{(+1)} \cos \left( \frac{\vartheta_{\mathrm{ext}}}{2} - \beta_m^{(+1)} \right) +  B_m^{(-1)} \cos \left( \frac{\vartheta_{\mathrm{ext}}}{2} + \beta_m^{(-1)} \right) .
    \label{eq:defSignalDBDThreePath}
\end{split}
\end{equation}
Due to the three-path nature, it features two additional beating terms compared with SBD.
The overlap of outer paths that correspond to the ideal situation arises in analogy to SBD and is defined as
\begin{equation}
    \int \psi_{m,-1,1}^\ast \psi_{m,1,-1} \diffd p = \frac 1 2A_m \exp{(\operatorname{i} \alpha_m)}.
    \label{eq:defOverlapDBDtwopath}
\end{equation}
The overlap of the central spurious path with the outer paths is given by
\begin{equation}
     \int \psi_{m,\mp 1,\pm 1}^\ast \psi_{m,0,0} \diffd p = \frac 1 2 B_m^{(\pm 1)} \exp{\left[\operatorname{i} \beta_m^{(\pm 1)}\right]},
     \label{eq:defOverlapDBDthreepath}
\end{equation}
where $B_m^{(\pm 1)}$ is the respective amplitude and $\beta_m^{(\pm1)}$ its phase, discussed in the section below.

For a vanishing population of the central path (e.g., caused by some type of blow-away scheme), only the first line of Eq.~\eqref{eq:defSignalDBDThreePath} with $P_{m,0,0}=0$ gives rise to the signal.
We observe in Fig.~\ref{fig:overlapAbs}\,(b) that the amplitudes $A_m$ display the typical velocity-selective behavior for long durations and loss to higher diffraction orders for short durations.
Moreover, the overlap in the outer exit ports is the same, i.e., $A_1 = A_{-1}$.
In analogy to SBD, we also observe $A_0 \approx A_1 + A_{-1}$, where deviations occur primarily in the Raman-Nath regime.
It already gives a first hint that the sum of both outer exit ports plays the role of the second exit port in SBD, as one would naively expect from a three-level system.

In the previous section we observed that the central path becomes important for long durations, due to an increase of velocity selectivity.
When we include it in the description, two additional terms arise that oscillate with $\vartheta_\text{ext} / 2$ and lead to a beating with $4 \pi$ periodicity, see Eq.~\eqref{eq:defSignalDBDThreePath}.
This periodicity has also been observed in Ref.~\cite{he_phase_2021}.
Even though there is a massive increase in $P_{m,0,0}$ for long durations, Fig.~\ref{fig:overlapAbs}\,(c)-(e) shows the the overlap between the spurious central path and the intended outer paths displays no such behavior.
It can be explained as follows: the population not diffracted by the first pulse in the interferometer stems from the outer wings of the momentum distribution.
When the diffracted parts of the resonant paths are brought to interference with the central path, they have different momenta.
Therefore, they do not overlap.
As a consequence, the central path predominantly acts as a background population $P_{m,0,0}$ to the exit ports and acts similar to an incoherent contribution.
As such, it leads to a loss of contrast.
To remove such a background, one could project on narrower momentum intervals defining the exit port, or apply some sort of blow-away scheme.

Even though much smaller than the background population, the overlap of the spurious path is not vanishing.
We observe in Fig.~\ref{fig:overlapAbs}\,(c) and~(d) that the amplitude of the beating in the outer exit ports is larger than the one in the central port, shown in Fig.~\ref{fig:overlapAbs}\,(e).
For the central exit port, the spurious amplitudes are small but symmetric, e.g., $B_0^{(+1)} = B_0^{(-1)}$.
This phenomenon does not occur in the outer exit ports, where the amplitudes displayed in Fig.~\ref{fig:overlapAbs}\,(d) differ slightly from those shown in Fig.~\ref{fig:overlapAbs}\,(c) for short durations.

We define in analogy to Eq.~\eqref{eq:defOverlapDBDthreepath} the overlap of higher-order paths with the resonant ones as
\begin{equation}
    \left| \int \psi_{0,\mp 2,\pm 2}^\ast \psi_{0,-n,n} \diffd p \right| = \frac 1 2 C_n^{(\pm)}
\end{equation}
where again $m = 0$ denotes the exit port and $n=0,\pm1,\pm2$.

\begin{figure}[h]
\includegraphics[width = \columnwidth]{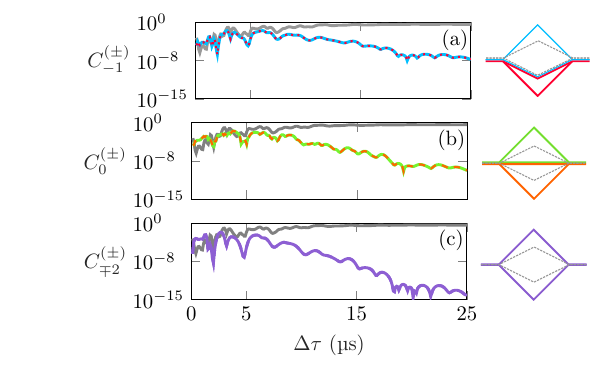}
\caption{
    Moduli of overlaps $ C_n^{(\pm)}$ including the higher-order paths $\psi_{0,\mp 2,\pm 2}$ and ending in exit port $0$ on a logarithmic scale and in the Raman-Nath regime.
    The diagrams on the right connect each plot to the corresponding two paths and exit port trough a color code.
    Panel (a) shows the overlap of the higher-order paths with desired ones, while panel (b) shows those with the central path.
    In contrast, panel (c) shows the overlap of both higher-order paths.
    The overlap of the two desired paths are drawn in gray for comparison.
    As expected, all overlaps including higher-order paths are magnitudes smaller outside the Raman-Nath regime.
    Thus, in the Bragg regime, their contribution can be neglected.
    Add sentence: "We used $\Delta \wp = 0.05 \hbar K$ and $T = 0.1 \mathrm{s}$." 
    }
\label{fig:overlapHigherOrder}
\end{figure}

Figure~\ref{fig:overlapHigherOrder} shows the results on a logarithmic scale where we focus on the Raman-Nath regime, i.e. short pulse durations.
We compare the overlaps of higher-order paths to the ones of the desired paths from Fig.~\ref{fig:overlapAbs}\,(b), which is displayed as a gray line.
The overlaps of higher-order paths with the desired ones are all of the same order, as can be seen in Fig.~\ref{fig:overlapHigherOrder}\,(a) and\,(b).
The overlaps between both higher-order paths is shown in Fig.~\ref{fig:overlapHigherOrder}\,(c) and orders of magnitude smaller, as expected from our discussion of Sec.~\ref{subsec:contributions}.
However, all those overlaps only contribute significantly for extremely short pulse durations, i.e., in the Raman-Nath regime.
Consequently, they can be ignored for an operation with pulses in the Bragg and quasi-Bragg regime as highlighted by the logarithmic plot.

\subsection{Influence of phase imperfection}
\label{subsec:phase}

\begin{figure}[h]
\includegraphics[width = \columnwidth]{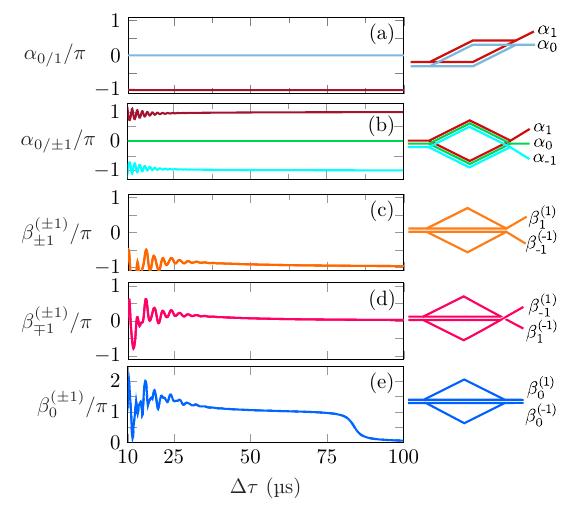}
\caption{
    Phase of $\alpha_m$ and $\beta_m^{(\pm1)}$ of the overlap in exit port $m$.
    They correspond to phase difference accumulated during the propagation along different paths.
    While $\alpha_m$ corresponds to the phase difference between two desired paths, $\beta_m^{(\pm)}$ is the phase difference between one spurious and one desired path.
    The diagrams on the right connects each phase difference to the corresponding two paths and exit port through a color code.
    Panel (a) shows that in SBD the phase difference between the two paths is independent of the duration of the pulse.
    Moreover, both ports are phase shifted by $\pi$ as expected.
    While one also observes in panel (b) an expected phase shift of $\pi$ between the central exit port and the outer ones in DBD in the Bragg regime, additional but symmetric phases arise in the Raman-Nath regime. 
    The phases that arise from an interference with the spurious central path are shown in panels (c)-(e) and show that additional phase contributions have to be taken into account in the beating contributions.
    For these simulations we used an initial Gaussian momentum width of $\Delta \wp = 0.05 \hbar K$.
}
\label{fig:overlapPhase}
\end{figure}

So far, we regarded only the modulus of the overlap to describe visibility and beating.
However, the phase induced by imperfect mirrors and beam splitters may also influence the signal.
For the case of SBD, it is encoded into $\alpha_m$.
However, as we see in Fig.~\ref{fig:overlapPhase}\,(a), the phase of the overlap in the exit ports $0$ and $1$ is independent of the regime. Moreover, the phase between the exit ports is only shifted by $\alpha_0-\alpha_1 \cong \pi$ as expected from an ideal two-level system.
In fact, no spurious phase is observed in a SBD Mach-Zehnder interferometer that arises from beam-splitter imperfections. 

\begin{figure}[h]
\includegraphics[width = \columnwidth]{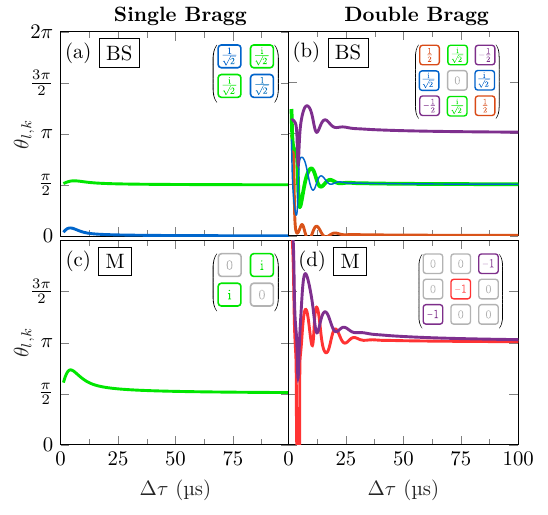}
\caption{
    Phase $\theta_{l,k}$ of the transition function $G_{l,k}(p = 0)$.
    The top row [panels (a) and (b)] shows  the results of the relevant transition elements of a beam splitter, while the bottom row [panels (c) and (d)] shows the results for a mirror pulse.
    The colors can be identified with the matrix elements for resonant Rabi oscillations of a two-level system in SBD given in the top right corners, which corresponds to the elements $G_{l,k}(0)$ with $l,k=0,1$.
    For DBD the colors are identified with the matrix elements for generalized resonant Rabi oscillations of a three-level system, which corresponds to the elements $G_{l,k}(0)$ with $l,k=-1,0,1$.
    We observe for large pulse durations, i.\,e. in the deep Bragg regime, that the phases converge to those expected from resonant Rabi oscillations in a two- or three-level system, respectively.
    There are, however, deviations in the Raman-Nath regime.
}
\label{fig:diffractionPhase}
\end{figure}

To explain this observation, we study the phase imprinted by the individual diffracting elements.
For that, we consider the phase
\begin{equation}
	 \theta_{l,k} \equiv \mathrm{arg} \{ G_{l,k}(p = 0) \}
\end{equation}
of the transition function $G_{l,k}$ for $p = 0$.
For a SBD beam splitter, the phases are plotted in Fig.~\ref{fig:diffractionPhase}\,(a), for a mirror in Fig.~\ref{fig:diffractionPhase}\,(c).
The elements of an ideal beam splitter and mirror matrix are shown in the top right of each panel and are encoded by the same color as the phase extracted from the numerical simulations.
For long pulse durations (i.e., in the Bragg regime), we recover the ideal phases from a perfect two-level system and observe the phase $\pi / 2$ when diffracted.
In the Raman-Nath regime, we see deviations.
However, the phases of the off-diagonal elements always agree and so do the ones of the diagonal elements.
For exit port $0$ of a Mach-Zehnder interferometer, the upper path is diffracted by the lower left matrix element and by the mirror.
The lower path is diffracted by the mirror and by the upper right matrix element. 
Because the same phases are imprinted on the off diagonal, both paths acquire the same spurious phase so that it cancels in the interference signal.
A similar observation is made for $p\neq 0$, so that averaging over all momenta does not lead to an additional phase or loss of contrast.

In DBD, the phases of the matrix elements in Fig.~\ref{fig:diffractionPhase}\,(b) and (d) still converge to those of a perfect three-level system in the Bragg regime and display some symmetry.
On the upper path, a blue beam splitter, a purple mirror and a green beam splitter diffract into the central exit port.
The same is true for the lower branch.
As a result, we expect no observed spurious phase shift in the interference signal, which coincides with $\alpha_0=0$ in all regimes from the numerical simulation of the overlap in Fig.~\ref{fig:overlapPhase}\,(b).

However, in exit port $+1$, the upper branch is diffracted by a blue beam splitter, a purple mirror and a purple beam splitter again, whereas the lower branch is diffracted by a blue beam splitter, a purple mirror and an orange beam splitter.
In this case, spurious phases are not imprinted symmetrically anymore, which can be observed in Fig.~\ref{fig:overlapPhase}\,(b).
Consequently, the influence of the spurious phase should reflect itself in the outer exit port.
For the exit port $-1$ the role of the last beam splitter is inverted, so we expect an opposite phase in complete agreement with Fig.~\ref{fig:overlapPhase}\,(b).
In fact, we find $\alpha_{+1} = -\alpha_{-1}$ , which implies that the interference patterns of the outer exit ports are shifted in opposite directions.
Moreover, we find $\alpha_{+1}-\alpha_{-1} \cong 2 \pi$ in the Bragg regime which shows that the average differential phase of the outer exit port is shifted by $\pi$ with respect to the central one.

For the spurious paths and the three-path interference pattern, similar considerations apply.
Whereas the central path is symmetric with $\beta^{(+1)}_0 = \beta^{(-1)}_0 $ (similar to $\alpha_0=0$) as shown in Fig.~\ref{fig:overlapPhase}\,(e), for other parts we observe the contribution $\beta^{(\pm 1)}_{+1} - \beta^{(\pm1)}_{-1} \cong \pi$ in the Bragg regime, which is plotted in Fig.~\ref{fig:overlapPhase}\,(c) and (d).

\subsection{Mach-Zehnder interference signal}
\label{subsec:signal}

With the discussion of the amplitude and the phase of the overlap, we are in the position to describe the whole interference signal generated by resonant first-order diffraction.
For SBD, only two paths contribute to the signal and no beating arises.
Moreover, the observed pattern experiences no phase shift caused by imperfect diffraction.
Even though loss caused by velocity selectivity in the Bragg regime leads to a decreased number of detected particles, the contrast is not affected and remains almost perfect.
However, in DBD the background stemming from the central path leads to a degrading contrast.

\begin{figure}[h]
\includegraphics[width = \columnwidth]{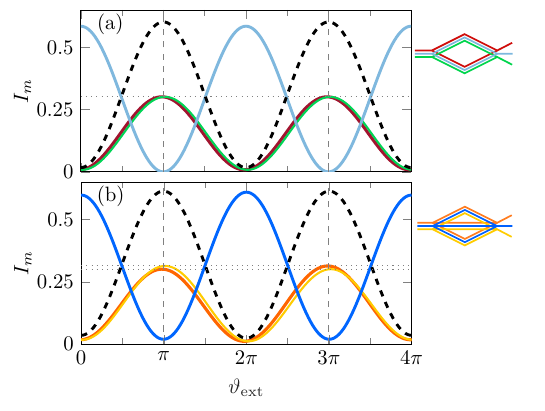}
\caption{
    Interference signal of a DBD Mach-Zehnder interferometer, where an external phase $\pm\vartheta_\text{ext}/2$ was imprinted symmetrically on both resonant arms.
    The legends on the top right show the respective color-coded paths that interfere in each exit port, the dashed line shows the sum of the two outer ports.
    Panel (a) displays the signal that arises from two-path interference where we assume $\psi_{0,0,0}=0$.
    We observe in all exit ports perfect visibility.
    Moreover, there is a phase shift between the outer exit ports which, if the populations are added, cancels out.
    Panel (b) shows the signal for three-path interference of DBD.
    Due to the spurious central path, we observe a loss of contrast.
    The main contribution arises from the background population of the central path ending in the central exit port.
    The nonvanishing overlap gives rise to an additional beating and a $4\pi$ periodicity, which can be observed more prominently in the outer two exit ports.
    The horizontal dotted lines indicate the different amplitudes of this periodicity.
    For these simulations we used $\Delta \tau = \SI{62.5}{\micro \second}$, $\Delta \wp = 0.05 \hbar K$, and an interferometer time of $T = \SI{0.1}{\second}$.
    }
\label{fig:Signal}
\end{figure}

In Fig.~\ref{fig:Signal} we display the interference signal as a function of the external phase $\vartheta_{\mathrm{ext}} \in \{ 0,4\pi \}$.
Panel~(a) shows two-path interference, for example realized by blow-away pulses with  $B_m^{(\pm 1)}= B_0 = 0=P_{m,0,0}$.
As expected from $\alpha_0=0$, the pattern $I_0$ shows no spurious phase shift.
However, the patterns of the outer ports $I_{\pm 1}$ are shifted in opposite direction as implied by $\alpha_{+1}+\alpha_{-1} = 0 $.
Because in addition the amplitudes are the same, the sum $I_{+1}+ I_{-1}$ is of opposite phase to $I_0$ and exhibits no spurious phase shifts.
Indeed, from Eq.~\eqref{eq:defSignalDBDThreePath} we find the relation
\begin{align}
\begin{split}
    I_1+I_{-1} =&  \sum\limits_{m=\pm1}(P_{m,-1,1} + P_{m,0,0} + P_{m,1,-1})\\
    &+ (A_1+A_{-1}) \cos\frac{\alpha_1-\alpha_{-1}}{2}  \cos \left( \vartheta_{\mathrm{ext}} - \frac{\alpha_1+\alpha_{-1}}{2} \right)\\
    &+(A_{-1}-A_1) \sin\frac{\alpha_1-\alpha_{-1}}{2}  \sin \left( \vartheta_{\mathrm{ext}} - \frac{\alpha_1+\alpha_{-1}}{2} \right)
\end{split}
\label{eq:sumSignalsOuterPorts}
\end{align}
where the trigonometric identities for the sum and difference of cosine functions have been used and all contributions proportional to $B_m^{(\pm 1)}=0$ have been neglected.
Moreover, the phase correction of the remaining cosine term vanishes because $\alpha_{+1} + \alpha_{-1} = 0$ in the Bragg regime.
Similarly, the prefactor that is a cosine reduces to $-1$ in the regime, where $\alpha_{+1} - \alpha_{-1} \cong 2 \pi$.

For three-path interference, the situation is more subtle, as shown in Fig.~\ref{fig:Signal}\,(b).
In addition to the effects that occur for two-path interference, the interference patterns exhibit a beating with phase $\vartheta_\text{ext}/2$ and a $4 \pi$ periodicity, as implied by Eq.~\eqref{eq:defSignalDBDThreePath}.
Because the overlap is larger for the outer exit ports, we observe a more prominent beating in $I_{\pm1}$.
However, based on symmetries of phases and amplitudes, the beating of the sum $I_{-1} + I_{+1}$ is suppressed in the Bragg regime, where also no spurious phase shifts occur, similar to the interference signal in $I_0$.
This feature underlines the benefits of treating $I_{-1} + I_{+1}$ as a joint exit port, as expected in a dressed state picture for three-level systems \cite{radmore_population_1982}.
It is one of the reasons why both exit ports were added in the experimental implementation of double Bragg interferometry~\cite{Ahlers2016}.

\subsection{Effects of Doppler detuning}

\begin{figure}
\includegraphics[width = \columnwidth]{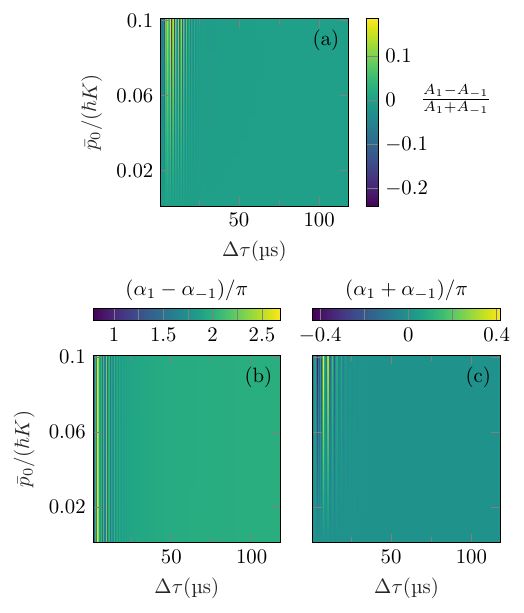}
\caption{
    Effects of Doppler detuning on the sum of the DBD interference patterns of both outer exit ports.
    In principle a richer structure in the pattern arises for a nonvanshing difference $A_1 - A_{-1}$, see Eq.~\eqref{eq:sumSignalsOuterPorts}.
    We display this difference in panel (a) and observe that it is primarily dominated by the duration and vanishes in the Bragg regime, as expected.
    Similarly, panel (b) shows the phase difference $\alpha_1 - \alpha_{-1}$ of both exit ports that leads to a loss of contrast.
    It only differs from $2\pi$ in the Raman-Nath regime and does not exihit a big dependence on the initial momentum.
    The sum $\alpha_1 + \alpha_{-1}$ is plotted in panel (c), which shifts the interference fringe.
    We also observe a dependence on the Doppler detuing, but only for short durations, while outside this regime it rapidly vanishes as discussed in the section above.
    As a consequence, we expect no further effects from imperfect pulses in the regime usually used for atom interferometry. 
    In these simulations we used $T = \SI{0.1}{\second}$ and $\Delta \wp = 0.05 \hbar K$.
}
\label{fig:signalPhases}
\end{figure}

An initial momentum $\bar p_0$ introduces a Doppler shift that acts similar to a detuning of diffraction processes.
Because of its symmetry, DBD has to be performed with ideally vanishing initial velocity.
Hence, we focus only on small values that deviate from this resonance and disturb the process.
Such values can be a consequence of imperfect preparation and release in a microgravity environment or of a misalignment of horizontal configurations on ground.
Such a Doppler detuning can be mitigated in SBD by adjusting the laser frequencies.
However, this treatment is not possible for DBD using counterpropagating gratings in a retroreflective geometry, because both lattices are accelerated in opposite directions.
We therefore study the effect of nonvanishing initial  $\bar p_0$ in DBD and whether the symmetries identified in the sections above still persist in Fig.~\ref{fig:signalPhases}, where we show the results of our study. 

In fact, we observe in Fig.~\ref{fig:signalPhases}\,(a) $A_1 \neq A_{-1}$ in the Raman-Nath regime and for increasing $\bar p_0$, so that in principle the third line of Eq.~\eqref{eq:sumSignalsOuterPorts} contributes and gives rise to a richer interference pattern.
However, there is no observable effect in the Bragg regime.
This additional contribution is further suppressed for small $\alpha_1 - \alpha_{-1}$.
We plot this phase difference in Fig. ~\ref{fig:signalPhases}\,(b) and observe that it is in fact small, with only contributions that arise in the the Raman-Nath regime.
While it is a dependence on the Doppler detuning, the effect only arises for short pulse durations outside the Bragg regime.
The phase error $\alpha_1 + \alpha_{-1}$ that shifts a fringe even without beating is shown in Fig. ~\ref{fig:signalPhases}\,(c).
Similar to the phase difference, it is primarily a feature of the regime rather than the Doppler detuning, although one observes a dependence on $\bar p_0$.
As a consequence, the sum of both exit ports shows no significant beating or phase errors even for Doppler-detuned DBD in the Bragg regime.

\section{Influence of gravity}
\label{sec:gravity}

To analyze the influence of gravity, we introduce a gravitational acceleration $a$ into the simulations that has three effects:
(i) It introduces a phase difference between the arms of the interferometer that is measured by gravimeters and included in Eqs.~\eqref{eq:kgtsquaredsingle} and~\eqref{eq:kgtsquareddouble}.
(ii) Between pulses, the atoms gain momentum that ideally has to be compensated for by adjusting the resonance condition.
(iii) The atoms accelerate during the pulses and drop out of resonance, which leads to a decreased diffraction efficiency.

In applications with SBD, deleterious effects can be mitigated by a frequency chirp \cite{mcguirk_sensitive_2002}.
This technique also serves to read out gravity by finding the zero fringe as a function of the chirping rate and locking it to the acceleration \cite{peters_high-precision_2001}.
When the diffraction lasers are chirped, their instantaneous frequency difference is adjusted to match the resonance condition.
In practice, the instantaneous frequency difference $\dot \phi(t)$ between both light fields takes the form
\begin{equation}
\label{eq:inst_freq_diff}
    \dot \phi(t) = \omega_b(t) - \omega_a(t) = \Delta \omega + \gamma t
\end{equation}
where $\gamma$ denotes the linear chirping rate and $\Delta \omega$ the frequency difference that encodes resonant diffraction at $t = 0$.

Integrating Eq.~\eqref{eq:inst_freq_diff}, we obtain the time-dependent laser phase difference
\begin{equation}
\phi(t) = \phi_0 + \Delta \omega t + \gamma t^2 /2
\end{equation}
with an offset laser phase $\phi_0$.
We include this laser phase in the set of differential equations provided in Appendix~\ref{app:diffODE}.
They are given in an interaction picture that encodes the acceleration of the atoms during the pulse as a time-dependent exponential.
It is straightforward to see that for $\gamma = - K a$ the effects of gravity cancel.

Alternatively, the close relation between the phases due to gravitational acceleration and to the frequency chirp can be understood by considering the transformation from the laboratory frame to a freely falling one \cite{Roura2020}.
Indeed, in a freely falling frame the effect of gravitational acceleration results in a frequency chirp rate $K a$.
Therefore, the associated laser phases can be compensated with a suitable chirp rate of the injected laser frequencies $\gamma = -K a$.
In practice, however, $\gamma$ will only coincide approximately with $-K a$ and some residual phase contributions may remain.

In contrast to SBD, the effects of gravity cannot be mitigated in DBD when the two counterpropagating gratings are connected by retroreflection because chirping the frequency difference will accelerate them in opposite directions (so that only one of the two diffraction processes remains resonant while the one in the opposite direction becomes increasingly Doppler detuned).
Therefore, its use is restricted to microgravity conditions or nearly horizontal configurations where small angles of the order of microradians lead to a projected gravitational acceleration of order $10^{-5}\ \mathrm{m/s}^2$ or less~%
\footnote{
In both cases it is possible to have SBD rather than DBD, even for vanishing initial velocities, by simultaneously chirping both injected frequencies at a sufficiently high rate and exploiting the extra time of flight for the retroreflected components \cite{Perrin2019}.
Alternatively, it is also possible to avoid DBD by considering magnetically sensitive states and suitable laser polarizations \cite{Bernard2022}.
}.
On the other hand, the effects of gravity can be compensated in interferometers based on DBD and employing retroreflected laser beams by injecting a third frequency component \cite{Malossi_Double_2010}, but then laser phase noise is no longer entirely common to both interferometer arms.
Interestingly, through a suitable choice of the three injected frequencies the contribution of laser phase noise cancels out in a differential acceleration measurement for two different atomic species \cite{zhou_test_2015}.

To minimize deleterious effects of gravity, we study a setup similar to an atomic fountain, where the mirror pulse is resonant at the apex, even though we work with much smaller accelerations than conventional fountain experiments.
For that, we chose the initial momentum of the wave function such that it compensates the drop due to gravity and is on resonance at the center of the mirror pulse.
In this case both beam splitters are off-resonant but symmetric.
Our treatment could be also applied to other initial conditions more common for small accelerations.

\begin{figure}
\includegraphics[width = \columnwidth]{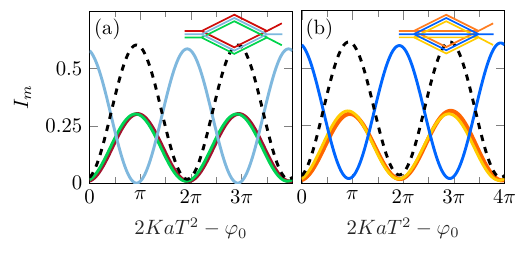}\textbf{}
\caption{
    Interference signal of a DBD Mach-Zehnder interferometer induced by gravity.
    The legends on the top right show the respective color-coded paths that interfere in each exit port, the dashed line shows the sum of the outer two ports.
    Panel (a) displays the signal that arises from two-path interference, while panel (b) shows the signal for three-path interference of DBD.
    The phase is varied by changing the interferometer time $T$ and the offset phase is $\phi_0=50\pi$ which corresponds to $T = \SI{0.60}{\second}$.
    We observe a qualitative behavior similar to Fig.~\ref{fig:Signal}.
    For these simulations we used $\Delta \tau = \SI{62.5}{\micro \second}$, $a = \SI{10}{\micro \metre \per \square \second}$ and $\Delta \wp = 0.05 \hbar K$.
}
\label{fig:signalGravity}
\end{figure}

In this setup, we perform DBD simulations for different interrogation times $T$ in microgravity conditions or horizontal configurations, where we use a projected gravitational acceleration of \SI{10}{\micro \metre \per \square \second}.
This way, we scan the interferometer phase $\varphi_0 + \vartheta_g= 2 K a T^2$ over a $4\pi$ interval with an offset of $\varphi_0 = 50 \pi$. 
The role of the external phase is now played by gravity and included in our simulations.
Using the same derivation of Eqs.~\eqref{eq:defSignalDBDTwoPath} and~\eqref{eq:defSignalDBDThreePath} with $\vartheta_\text{ext}=0$, we observe similar features as before:
Two-path interference induces a small phase shift between the outer exit ports as observed in Fig.~\ref{fig:signalGravity}\,(a) and
three-path interference induces an offset of decreased visibility as well as a beating as seen in Fig.~\ref{fig:signalGravity}\,(b).

As discussed above, the two-path interference pattern $I_0$ shown in Fig.~\ref{fig:signalGravity}\,(a) exhibits no beating, but of course depends on the gravity- and pulse-duration dependent phase shift.
The sum of the ports $I_{1} + I_{-1}$ exhibits an analogous behavior.

Gravity breaks the symmetries pointed out above, so that additional phase shifts during diffraction arise.
These effects are expected for two-level systems, where residual acceleration during a pulse deteriorates the quality of the Rabi oscillations~\cite{lammerzahl_rabi_1995,marzlin_freely_1996}.
The influence of finite pulse durations has been studied in the context of single Raman diffraction~\cite{bertoldi_phase_2019,antoine_matter_2006}, but but can be transferred to SBD assuming that higher-order diffraction can be neglected.

\begin{figure}
\includegraphics[width = \columnwidth]{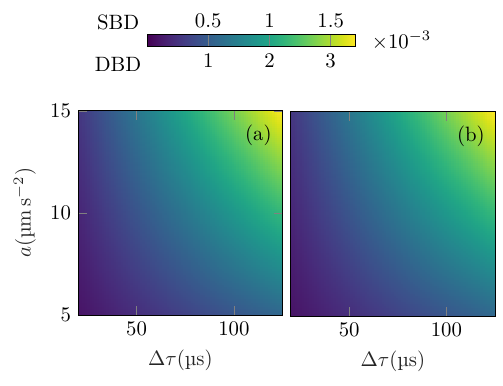}
\caption{
    Phase difference between the two desired paths in the central exit port of DBD caused by the finite pulse duration and gravity, and compared with SBD.
    In panel (a) we plot $\alpha_0-K a T^2$ for SBD, while panel (b) shows $\alpha_0-2K a T^2$ for DBD.
    We see that the phases differ by a factor of two in the Bragg regime.
    The scale of the color coding of panel (a) is indicated above the color bar, whereas the scale of panel (b) is indicated below the color bar.
    For these simulations, we used an initial Gaussian momentum width of $\Delta \wp = 0.01 \hbar K$ and an interferometer time of $T = \SI{0.01}{\second}$.
}
\label{fig:alphaZero}
\end{figure}

In contrast to the case of box-shaped pulses with analytical expressions, we numerically study Gaussian pulses and start with a discussion of SBD.
For that, we show in Fig.~\ref{fig:alphaZero}\,(a) the phase difference $\alpha_0-K a T^2$ acquired during the pulse as a function of $\Delta \tau$ and $a$.
The phase $\alpha_0$ is calculated by subtracting the phase imprinted on the upper path from the phase of the lower path, thus the positive sign of the gravitational phase.
As expected, for higher pulse durations and gravity, the phase shift increases.
For short pulses in the Raman-Nath regime, diffraction cannot be described by a two-level system any more.
However, since in this regime the atoms do not have the time to experience a considerable acceleration, all effects of gravity are suppressed.
Moreover, we always observe a phase difference $\alpha_1 - \alpha_{0} = \pi$ for SBD, such that both exit ports are opposite in phase, taking finite pulse durations into account.

For DBD, the phase difference $\alpha_0-2K a T^2$ is associated with phases acquired for finite pulse durations between both desired paths.
It behaves similar to the single Bragg case, see Fig.~\ref{fig:alphaZero}\,(b), and differs exactly by a factor two in the Bragg regime.
This behavior can be attributed to doubled momentum transfer.

As mentioned above and shown in Fig.~\ref{fig:signalGravity}\,(b), a beating arises for three-path interference.
A Fourier analysis of the beating pattern of $I_{+1} + I_{-1}$ (and $I_0$) reveals that it consists of two components.
The first one originates from the two desired paths and is observed in two-path interference.
The other component is suppressed and differs by a factor of one half, in complete analogy to the external phase of Sec.~\ref{subsec:signal}.
A similar analysis can be performed in experimental applications.

\section{Conclusions}
\label{sec:conclusion}

We have analyzed the contributions of interferometer paths to specific exit ports of a  Mach-Zehnder interferometer based on first-order Bragg diffraction and concluded that higher-order contributions are not relevant in SBD if spurious ports can be spatially resolved at detection.
In contrast, for DBD the central path always contributes.
Even though atom-optical phase imperfections also arise in SBD, they cancel out in a Mach-Zehnder interferometer due to its intrinsic symmetry.
The only surviving phases stem from a finite pulse duration under the influence of gravity.
In DBD on the other hand, the non-negligible spurious central path gives rise to a beating pattern due to three-path interference.
The three relevant momentum states lead to a spurious phase shift between the outer exit ports caused by phase errors.
We demonstrated that the sum of their populations gives rise to a signal without such deleterious effects.

For the Mach-Zehnder interferometers we analyzed the spurious central path that can lead to contrast reduction in DBD, whereas no loss of contrast arises in SBD (provided that the effects of gravity are entirely compensated through a suitable chirp of the laser frequency difference).
Nevertheless, in actual experiments a reduction of interferometer contrast can also be caused by other effects such as technical detection noise, wavefront distortions and finite beam size.
In addition, rotations and gravity gradients lead to relative displacements between the interfering wave packets and result in contrast losses that can be particularly significant for long interferometer times \cite{Roura2014}.
However, these losses can be avoided thanks to effective mitigation techniques existing for both rotations \cite{Hogan2009,Lan2012,Dickerson2013} and gravity gradients \cite{Roura2014,Roura2017,Overstreet2018}. 

The approach can in principle be applied to double Raman diffraction too, where the Raman-Nath regime does not show such a rich structure~\cite{Hartmann2020a}, but higher-order diffraction is possible as well~\cite{hartmann_atomic_2020}.
Furthermore, since the cancellation of diffraction phases was caused by the symmetry of the Mach-Zehnder interferometer, it might be worthwhile to check whether other geometries like Ramsey-Bord{\'e}-type setups display similar symmetries~\cite{estey_high-resolution_2015}.
Possible options to mitigate diffraction phases in DBD could be double-diamond (or butterfly) geometries \cite{schubert_multi-loop_2021,sidorenkov_tailoring_2020} with two central $\pi$ pulses, which are however insensitive to time-dependent linear accelerations.
Here, we have considered perfect plane waves as diffracting beams, but effects from beam shapes lead to additional contributions \cite{neumann_aberrations_2021,mielec_atom_2018,trimeche_concept_2019}.

Throughout this article, we focused on resonant first-order diffraction.
However, resonant higher-order diffraction \cite{siems_analytic_2020,Ahlers2016,Szigeti_2012,altin_precision_2013} leads to more spurious interferometer paths that affect the signal as well \cite{beguin_characterization_2021}.
We hope to stimulate further studies which focus on higher diffraction orders.
Like off-resonant effects for first order, such off-resonant diffraction should be suppressed and a discussion similar to the one above seems possible.
However, resonant higher-order diffraction requires an operation in quasi-Bragg regime \cite{muller_atom-wave_2008} with fairly short pulse durations, which means that off-resonant effects become in general more relevant.
Extending our study to resonant higher-order diffraction seems to be a necessary step for the analysis of large-momentum-transfer \cite{gebbe_twin-lattice_2021,muller_atom_2008,chiow_102hbark_2011,mcdonald80bark2013} atom interferometry and can, in principle, be performed with the techniques presented in this article.

\begin{acknowledgements}
We are grateful to W. P. Schleich for his stimulating input and continuing support.
We thank the QUANTUS and INTENTAS teams for fruitful and interesting discussions.
The work of IQ\textsuperscript{ST} is financially supported by the Ministry of Science, Research and Art Baden-W\"urttemberg (Ministerium f\"ur Wissenschaft, Forschung und Kunst Baden-W\"urttemberg).
The QUANTUS and INTENTAS projects are supported by the German Aerospace Center (Deutsches Zentrum f\"ur Luft- und Raumfahrt, DLR) with funds provided by the Federal Ministry for Economic Affairs and Climate Action (Bundesministerium f\"ur Wirtschaft und Klimaschutz, BMWK) due to an enactment of the German Bundestag under Grant Nos. 50WM1956 (QUANTUS V), 50WM2250D-2250E (QUANTUS+), as well as Nos. 50WM2177-2178 (INTENTAS).
E.G. thanks the German Research Foundation (Deutsche Forschungsgemeinschaft, DFG) for a Mercator Fellowship within CRC 1227 (DQ-mat).
A.~R.\ is supported by the Q-GRAV Project within the Space Research and Technology Program of the German Aerospace Center (Deutsches Zentrum für Luft- und Raumfahrt, DLR).
\end{acknowledgements}

\appendix
\section{Calculation of transition function}
\label{app:diffODE}

We solve the effective Schr\"odinger equation in momentum representation which, after the rotating wave approximation~\cite{schleich_quantum_2001}, takes the form
\begin{widetext}
\begin{equation}
	\begin{split}
		 \frac{\text{d}\psi_n(p_0,t) }{\text{d}t} = \imunit \Omega(t-t_j) \Big[ &\mathrm e^{- \imunit \omega_\mathrm{D}  (t-t_j)} \mathrm e^{\imunit K a (t-t_j)^2 / 2} \mathrm e^{\imunit \phi(t)}  \mathrm e^{- \imunit (2n + 1) \omega_K (t-t_j)} \, \psi_{n+1}(p_0,t)  \\
		 & +\mathrm e^{\imunit \omega_\mathrm{D}  (t-t_j)} \mathrm e^{-\imunit K a (t-t_j)^2 / 2} \mathrm e^{-\imunit \phi(t)}  \mathrm e^{\imunit (2n - 1) \omega_K (t-t_j)} \, \psi_{n - 1}(p_0,t) \Big]
	\end{split}
	\label{eq:singleBraggODE}
\end{equation}
for SBD and 
\begin{equation}
	\begin{split}
        \frac{\text{d}\psi_n(p_0,t) }{\text{d}t} = & \imunit \Omega(t-t_j) \, \mathrm e^{\mathrm{i} \omega_\mathrm{D} (t-t_j)} \mathrm e^{-\imunit \frac{1}{2}K a (t-t_j)^2}  \left[ \mathrm e^{\imunit \phi(t)} \mathrm e^{ \imunit (2n - 1) \omega_K (t-t_j)} + \mathrm e^{-\imunit \phi(t)} \mathrm e^{\imunit (2n - 1) \omega_K (t-t_j)} \right]\psi_{n - 1}(p_0,t)\\
		 + &\imunit \Omega(t-t_j) \, \mathrm e^{-\mathrm{i} \omega_\mathrm{D} (t-t_j)} \mathrm e^{\imunit \frac{1}{2} K a (t-t_j)^2}   \left[ \mathrm e^{-\imunit \phi(t)} \mathrm e^{ -\imunit (2n + 1) \omega_K (t-t_j)} + \mathrm e^{\imunit \phi(t)} \mathrm e^{-\imunit (2n + 1) \omega_K (t-t_j)} \right] \psi_{n+1}(p_0,t)
	\end{split}
	\label{eq:doubleBraggODE}
\end{equation}
\end{widetext}
for DBD~\cite{giese2013double}.
It emerges after the adiabatic elimination of an auxiliary state~\cite{brion_adiabatic_2007,bernhardt_coherent_1981,marte_multiphoton_1992} and is derived in an interaction picture with respect to the center-of-mass motion.
The respective unitary transformation takes the form $\exp \left\lbrace - \imunit \left[ \hat p ^2/(2M)+M a \hat z  \right] (t - t_j)/ \hbar \right\rbrace$.
The starting time of the pulse is denoted by $t_j$.
Hence, each pulse is calculated in its own interaction picture that is initialized at this time.
Based on the momentum representation $\psi(p)$ of the center-of-mass wave function of the atom in the ground state, we have defined $\psi_n(p_0,t)=\psi(p_0+n \hbar K)$ and the Doppler frequency $\omega_\mathrm{D}= p_0 K /M$.
Hence, the different diffraction orders are denoted by the index $n$.

To obtain the transition function $G(p^\prime,p)$, we solve the set of differential equations numerically and for an initial condition $\psi_n(p_0,t_j)=1$ and rearrange the solutions in matrix form.
For the numerical solution, we resort to \textsc{MATLAB}'s \cite{MATLAB:2021} ODE45 which is a Runge-Kutta-type solver~\cite{shampine_matlab_1997}.
For the simulations in \cref{fig:Contributions,fig:overlapAbs,fig:diffractionPhase,fig:overlapPhase,fig:alphaZero} we introduce a cutoff for $n= \pm 9$, which corresponds to  $\pm 9.5 \hbar K$ in momentum space and divide the interval $[-\hbar K / 2, \hbar K / 2]$ in 501 equally-spaced grid points.
For the simulations in \cref{fig:Signal,fig:signalGravity} we introduce a cutoff for $n= \pm 4$, which corresponds to  $\pm 4.5 \hbar K$ in momentum space and divide the interval $[-\hbar K / 2, \hbar K / 2]$ in $201$ equally-spaced grid points.
The relative numerical propagation error of \textsc{MATLAB}'s ODE toolbox was $10^{-3}$ and the absolute error $10^{-6}$.

For all simulations, we use a Gaussian envelope of the pulse that gives rise to the time-dependent Rabi frequency $\Omega(t) = \Omega_0 \exp \big[-(t - 4 \Delta \tau)^2/(2 \Delta \tau^2) \big]$.
Here, $\Omega_0$ denotes the peak two-photon Rabi frequency.
The pulses are cut off at $\pm 4 \Delta \tau$.
A Gaussian is a smooth function that minimizes non-adiabatic effects stemming from rapid changes in frequency.
Also, it constitutes a compromise between short pulse duration and an adiabatic envelope.
The pulse area $\mathcal A$ is defined as
\begin{equation}
    \mathcal A = \int \operatorname d t \mathcal N \Omega(t)
\end{equation}
where $\mathcal N = 2$ for SBD and $\mathcal N = \sqrt{2}$ for DBD.
Beam splitters are generated via $\mathcal A = \pi / 2$ and mirrors as $\mathcal A = \pi$.

Since rubidium 87 is one of the most common species employed in atom interferometry, we use its values for our simulations~\cite{steck2001rubidium}.
To be precise, we use its mass $M = \SI{1.4432e-25}{\kilogram}$ and the wave number $K = \SI{1.6106e7}{\per \metre}$ associated with the D2 transition, resulting in the recoil (angular) frequency $\omega_K = 2 \pi \times \SI{15.0839}{\kilo \hertz}$.

\bibliography{bibfile}

\begin{thebibliography}{71}%
\makeatletter
\providecommand \@ifxundefined [1]{%
 \@ifx{#1\undefined}
}%
\providecommand \@ifnum [1]{%
 \ifnum #1\expandafter \@firstoftwo
 \else \expandafter \@secondoftwo
 \fi
}%
\providecommand \@ifx [1]{%
 \ifx #1\expandafter \@firstoftwo
 \else \expandafter \@secondoftwo
 \fi
}%
\providecommand \natexlab [1]{#1}%
\providecommand \enquote  [1]{``#1''}%
\providecommand \bibnamefont  [1]{#1}%
\providecommand \bibfnamefont [1]{#1}%
\providecommand \citenamefont [1]{#1}%
\providecommand \href@noop [0]{\@secondoftwo}%
\providecommand \href [0]{\begingroup \@sanitize@url \@href}%
\providecommand \@href[1]{\@@startlink{#1}\@@href}%
\providecommand \@@href[1]{\endgroup#1\@@endlink}%
\providecommand \@sanitize@url [0]{\catcode `\\12\catcode `\$12\catcode
  `\&12\catcode `\#12\catcode `\^12\catcode `\_12\catcode `\%12\relax}%
\providecommand \@@startlink[1]{}%
\providecommand \@@endlink[0]{}%
\providecommand \url  [0]{\begingroup\@sanitize@url \@url }%
\providecommand \@url [1]{\endgroup\@href {#1}{\urlprefix }}%
\providecommand \urlprefix  [0]{URL }%
\providecommand \Eprint [0]{\href }%
\providecommand \doibase [0]{https://doi.org/}%
\providecommand \selectlanguage [0]{\@gobble}%
\providecommand \bibinfo  [0]{\@secondoftwo}%
\providecommand \bibfield  [0]{\@secondoftwo}%
\providecommand \translation [1]{[#1]}%
\providecommand \BibitemOpen [0]{}%
\providecommand \bibitemStop [0]{}%
\providecommand \bibitemNoStop [0]{.\EOS\space}%
\providecommand \EOS [0]{\spacefactor3000\relax}%
\providecommand \BibitemShut  [1]{\csname bibitem#1\endcsname}%
\let\auto@bib@innerbib\@empty
\bibitem [{\citenamefont {Kasevich}\ and\ \citenamefont
  {Chu}(1991)}]{kasevich_atomic_1991}%
  \BibitemOpen
  \bibfield  {author} {\bibinfo {author} {\bibfnamefont {M.~A.}\ \bibnamefont
  {Kasevich}}\ and\ \bibinfo {author} {\bibfnamefont {S.}~\bibnamefont {Chu}},\
  }\bibfield  {title} {\bibinfo {title} {Atomic interferometry using stimulated
  {Raman} transitions},\ }\href {https://doi.org/10.1103/PhysRevLett.67.181}
  {\bibfield  {journal} {\bibinfo  {journal} {Phys. Rev. Lett.}\ }\textbf
  {\bibinfo {volume} {67}},\ \bibinfo {pages} {181} (\bibinfo {year}
  {1991})}\BibitemShut {NoStop}%
\bibitem [{\citenamefont {Kleinert}\ \emph {et~al.}(2015)\citenamefont
  {Kleinert}, \citenamefont {Kajari}, \citenamefont {Roura},\ and\
  \citenamefont {Schleich}}]{Kleinert2015}%
  \BibitemOpen
  \bibfield  {author} {\bibinfo {author} {\bibfnamefont {S.}~\bibnamefont
  {Kleinert}}, \bibinfo {author} {\bibfnamefont {E.}~\bibnamefont {Kajari}},
  \bibinfo {author} {\bibfnamefont {A.}~\bibnamefont {Roura}},\ and\ \bibinfo
  {author} {\bibfnamefont {W.~P.}\ \bibnamefont {Schleich}},\ }\bibfield
  {title} {\bibinfo {title} {Representation-free description of light-pulse
  atom interferometry including noninertial effects},\ }\href
  {http://dx.doi.org/10.1016/j.physrep.2015.09.004} {\bibfield  {journal}
  {\bibinfo  {journal} {Phys. Reports}\ }\textbf {\bibinfo {volume} {605}},\
  \bibinfo {pages} {1} (\bibinfo {year} {2015})}\BibitemShut {NoStop}%
\bibitem [{\citenamefont {Bongs}\ \emph {et~al.}(2019)\citenamefont {Bongs},
  \citenamefont {Holynski}, \citenamefont {Vovrosh}, \citenamefont {Bouyer},
  \citenamefont {Condon}, \citenamefont {Rasel}, \citenamefont {Schubert},
  \citenamefont {Schleich},\ and\ \citenamefont {Roura}}]{Bongs2019}%
  \BibitemOpen
  \bibfield  {author} {\bibinfo {author} {\bibfnamefont {K.}~\bibnamefont
  {Bongs}}, \bibinfo {author} {\bibfnamefont {M.}~\bibnamefont {Holynski}},
  \bibinfo {author} {\bibfnamefont {J.}~\bibnamefont {Vovrosh}}, \bibinfo
  {author} {\bibfnamefont {P.}~\bibnamefont {Bouyer}}, \bibinfo {author}
  {\bibfnamefont {G.}~\bibnamefont {Condon}}, \bibinfo {author} {\bibfnamefont
  {E.}~\bibnamefont {Rasel}}, \bibinfo {author} {\bibfnamefont
  {C.}~\bibnamefont {Schubert}}, \bibinfo {author} {\bibfnamefont {W.~P.}\
  \bibnamefont {Schleich}},\ and\ \bibinfo {author} {\bibfnamefont
  {A.}~\bibnamefont {Roura}},\ }\bibfield  {title} {\bibinfo {title} {Taking
  atom interferometric quantum sensors from the laboratory to real-world
  applications},\ }\href {https://doi.org/10.1038/s42254-019-0117-4} {\bibfield
   {journal} {\bibinfo  {journal} {Nature Rev. Phys.}\ }\textbf {\bibinfo
  {volume} {1}},\ \bibinfo {pages} {731} (\bibinfo {year} {2019})}\BibitemShut
  {NoStop}%
\bibitem [{\citenamefont {L\'ev\`eque}\ \emph {et~al.}(2009)\citenamefont
  {L\'ev\`eque}, \citenamefont {Gauguet}, \citenamefont {Michaud},
  \citenamefont {Pereira Dos~Santos},\ and\ \citenamefont
  {Landragin}}]{PhysRevLett.103.080405}%
  \BibitemOpen
  \bibfield  {author} {\bibinfo {author} {\bibfnamefont {T.}~\bibnamefont
  {L\'ev\`eque}}, \bibinfo {author} {\bibfnamefont {A.}~\bibnamefont
  {Gauguet}}, \bibinfo {author} {\bibfnamefont {F.}~\bibnamefont {Michaud}},
  \bibinfo {author} {\bibfnamefont {F.}~\bibnamefont {Pereira Dos~Santos}},\
  and\ \bibinfo {author} {\bibfnamefont {A.}~\bibnamefont {Landragin}},\
  }\bibfield  {title} {\bibinfo {title} {Enhancing the {A}rea of a {R}aman
  {A}tom {I}nterferometer {U}sing a {V}ersatile {D}ouble-{D}iffraction
  {T}echnique},\ }\href {https://doi.org/10.1103/PhysRevLett.103.080405}
  {\bibfield  {journal} {\bibinfo  {journal} {Phys. Rev. Lett.}\ }\textbf
  {\bibinfo {volume} {103}},\ \bibinfo {pages} {080405} (\bibinfo {year}
  {2009})}\BibitemShut {NoStop}%
\bibitem [{\citenamefont {Berg}\ \emph {et~al.}(2015)\citenamefont {Berg},
  \citenamefont {Abend}, \citenamefont {Tackmann}, \citenamefont {Schubert},
  \citenamefont {Giese}, \citenamefont {Schleich}, \citenamefont {Narducci},
  \citenamefont {Ertmer},\ and\ \citenamefont
  {Rasel}}]{PhysRevLett.114.063002}%
  \BibitemOpen
  \bibfield  {author} {\bibinfo {author} {\bibfnamefont {P.}~\bibnamefont
  {Berg}}, \bibinfo {author} {\bibfnamefont {S.}~\bibnamefont {Abend}},
  \bibinfo {author} {\bibfnamefont {G.}~\bibnamefont {Tackmann}}, \bibinfo
  {author} {\bibfnamefont {C.}~\bibnamefont {Schubert}}, \bibinfo {author}
  {\bibfnamefont {E.}~\bibnamefont {Giese}}, \bibinfo {author} {\bibfnamefont
  {W.~P.}\ \bibnamefont {Schleich}}, \bibinfo {author} {\bibfnamefont {F.~A.}\
  \bibnamefont {Narducci}}, \bibinfo {author} {\bibfnamefont {W.}~\bibnamefont
  {Ertmer}},\ and\ \bibinfo {author} {\bibfnamefont {E.~M.}\ \bibnamefont
  {Rasel}},\ }\bibfield  {title} {\bibinfo {title} {Composite-light-pulse
  technique for high-precision atom interferometry},\ }\href
  {https://doi.org/10.1103/PhysRevLett.114.063002} {\bibfield  {journal}
  {\bibinfo  {journal} {Phys. Rev. Lett.}\ }\textbf {\bibinfo {volume} {114}},\
  \bibinfo {pages} {063002} (\bibinfo {year} {2015})}\BibitemShut {NoStop}%
\bibitem [{\citenamefont {Kovachy}\ \emph {et~al.}(2015)\citenamefont
  {Kovachy}, \citenamefont {Asenbaum}, \citenamefont {Overstreet},
  \citenamefont {Donnelly}, \citenamefont {Dickerson}, \citenamefont
  {Sugarbaker}, \citenamefont {Hogan},\ and\ \citenamefont
  {Kasevich}}]{kovachy2015quantum}%
  \BibitemOpen
  \bibfield  {author} {\bibinfo {author} {\bibfnamefont {T.}~\bibnamefont
  {Kovachy}}, \bibinfo {author} {\bibfnamefont {P.}~\bibnamefont {Asenbaum}},
  \bibinfo {author} {\bibfnamefont {C.}~\bibnamefont {Overstreet}}, \bibinfo
  {author} {\bibfnamefont {C.~A.}\ \bibnamefont {Donnelly}}, \bibinfo {author}
  {\bibfnamefont {S.~M.}\ \bibnamefont {Dickerson}}, \bibinfo {author}
  {\bibfnamefont {A.}~\bibnamefont {Sugarbaker}}, \bibinfo {author}
  {\bibfnamefont {J.~M.}\ \bibnamefont {Hogan}},\ and\ \bibinfo {author}
  {\bibfnamefont {M.~A.}\ \bibnamefont {Kasevich}},\ }\bibfield  {title}
  {\bibinfo {title} {Quantum superposition at the half-metre scale},\ }\href
  {https://www.nature.com/articles/nature16155} {\bibfield  {journal} {\bibinfo
   {journal} {Nature}\ }\textbf {\bibinfo {volume} {528}},\ \bibinfo {pages}
  {530} (\bibinfo {year} {2015})}\BibitemShut {NoStop}%
\bibitem [{\citenamefont {Ahlers}\ \emph {et~al.}(2016)\citenamefont {Ahlers},
  \citenamefont {M\"untinga}, \citenamefont {Wenzlawski}, \citenamefont
  {Krutzik}, \citenamefont {Tackmann}, \citenamefont {Abend}, \citenamefont
  {Gaaloul}, \citenamefont {Giese}, \citenamefont {Roura}, \citenamefont
  {Kuhl}, \citenamefont {L\"ammerzahl}, \citenamefont {Peters}, \citenamefont
  {Windpassinger}, \citenamefont {Sengstock}, \citenamefont {Schleich},
  \citenamefont {Ertmer},\ and\ \citenamefont {Rasel}}]{Ahlers2016}%
  \BibitemOpen
  \bibfield  {author} {\bibinfo {author} {\bibfnamefont {H.}~\bibnamefont
  {Ahlers}}, \bibinfo {author} {\bibfnamefont {H.}~\bibnamefont {M\"untinga}},
  \bibinfo {author} {\bibfnamefont {A.}~\bibnamefont {Wenzlawski}}, \bibinfo
  {author} {\bibfnamefont {M.}~\bibnamefont {Krutzik}}, \bibinfo {author}
  {\bibfnamefont {G.}~\bibnamefont {Tackmann}}, \bibinfo {author}
  {\bibfnamefont {S.}~\bibnamefont {Abend}}, \bibinfo {author} {\bibfnamefont
  {N.}~\bibnamefont {Gaaloul}}, \bibinfo {author} {\bibfnamefont
  {E.}~\bibnamefont {Giese}}, \bibinfo {author} {\bibfnamefont
  {A.}~\bibnamefont {Roura}}, \bibinfo {author} {\bibfnamefont
  {R.}~\bibnamefont {Kuhl}}, \bibinfo {author} {\bibfnamefont {C.}~\bibnamefont
  {L\"ammerzahl}}, \bibinfo {author} {\bibfnamefont {A.}~\bibnamefont
  {Peters}}, \bibinfo {author} {\bibfnamefont {P.}~\bibnamefont
  {Windpassinger}}, \bibinfo {author} {\bibfnamefont {K.}~\bibnamefont
  {Sengstock}}, \bibinfo {author} {\bibfnamefont {W.~P.}\ \bibnamefont
  {Schleich}}, \bibinfo {author} {\bibfnamefont {W.}~\bibnamefont {Ertmer}},\
  and\ \bibinfo {author} {\bibfnamefont {E.~M.}\ \bibnamefont {Rasel}},\
  }\bibfield  {title} {\bibinfo {title} {Double {B}ragg {I}nterferometry},\
  }\href {https://doi.org/10.1103/PhysRevLett.116.173601} {\bibfield  {journal}
  {\bibinfo  {journal} {Phys. Rev. Lett.}\ }\textbf {\bibinfo {volume} {116}},\
  \bibinfo {pages} {173601} (\bibinfo {year} {2016})}\BibitemShut {NoStop}%
\bibitem [{\citenamefont {Küber}\ \emph {et~al.}(2016)\citenamefont {Küber},
  \citenamefont {Schmaltz},\ and\ \citenamefont
  {Birkl}}]{kuber_experimental_2016}%
  \BibitemOpen
  \bibfield  {author} {\bibinfo {author} {\bibfnamefont {J.}~\bibnamefont
  {Küber}}, \bibinfo {author} {\bibfnamefont {F.}~\bibnamefont {Schmaltz}},\
  and\ \bibinfo {author} {\bibfnamefont {G.}~\bibnamefont {Birkl}},\ }\bibfield
   {title} {\bibinfo {title} {Experimental realization of double {Bragg}
  diffraction: robust beamsplitters, mirrors, and interferometers for
  {Bose-Einstein} condensates},\ }\href {http://arxiv.org/abs/1603.08826}
  {\bibfield  {journal} {\bibinfo  {journal} {\texttt{{arXiv}:1603.08826}}\ }
  (\bibinfo {year} {2016})}\BibitemShut {NoStop}%
\bibitem [{\citenamefont {Giese}\ \emph {et~al.}(2013)\citenamefont {Giese},
  \citenamefont {Roura}, \citenamefont {Tackmann}, \citenamefont {Rasel},\ and\
  \citenamefont {Schleich}}]{giese2013double}%
  \BibitemOpen
  \bibfield  {author} {\bibinfo {author} {\bibfnamefont {E.}~\bibnamefont
  {Giese}}, \bibinfo {author} {\bibfnamefont {A.}~\bibnamefont {Roura}},
  \bibinfo {author} {\bibfnamefont {G.}~\bibnamefont {Tackmann}}, \bibinfo
  {author} {\bibfnamefont {E.~M.}\ \bibnamefont {Rasel}},\ and\ \bibinfo
  {author} {\bibfnamefont {W.~P.}\ \bibnamefont {Schleich}},\ }\bibfield
  {title} {\bibinfo {title} {Double {B}ragg diffraction: {A} tool for atom
  optics},\ }\href {https://doi.org/10.1103/PhysRevA.88.053608} {\bibfield
  {journal} {\bibinfo  {journal} {Phys. Rev. A}\ }\textbf {\bibinfo {volume}
  {88}},\ \bibinfo {pages} {053608} (\bibinfo {year} {2013})}\BibitemShut
  {NoStop}%
\bibitem [{\citenamefont {Müller}\ \emph
  {et~al.}(2008{\natexlab{a}})\citenamefont {Müller}, \citenamefont {Chiow},
  \citenamefont {Long}, \citenamefont {Herrmann},\ and\ \citenamefont
  {Chu}}]{muller_atom_2008}%
  \BibitemOpen
  \bibfield  {author} {\bibinfo {author} {\bibfnamefont {H.}~\bibnamefont
  {Müller}}, \bibinfo {author} {\bibfnamefont {S.-w.}\ \bibnamefont {Chiow}},
  \bibinfo {author} {\bibfnamefont {Q.}~\bibnamefont {Long}}, \bibinfo {author}
  {\bibfnamefont {S.}~\bibnamefont {Herrmann}},\ and\ \bibinfo {author}
  {\bibfnamefont {S.}~\bibnamefont {Chu}},\ }\bibfield  {title} {\bibinfo
  {title} {Atom {Interferometry} with up to 24-{Photon}-{Momentum}-{Transfer}
  {Beam} {Splitters}},\ }\href {https://doi.org/10.1103/PhysRevLett.100.180405}
  {\bibfield  {journal} {\bibinfo  {journal} {Phys. Rev. Lett.}\ }\textbf
  {\bibinfo {volume} {100}},\ \bibinfo {pages} {180405} (\bibinfo {year}
  {2008}{\natexlab{a}})}\BibitemShut {NoStop}%
\bibitem [{\citenamefont {Siemß}\ \emph {et~al.}(2020)\citenamefont {Siemß},
  \citenamefont {Fitzek}, \citenamefont {Abend}, \citenamefont {Rasel},
  \citenamefont {Gaaloul},\ and\ \citenamefont
  {Hammerer}}]{siems_analytic_2020}%
  \BibitemOpen
  \bibfield  {author} {\bibinfo {author} {\bibfnamefont {J.-N.}\ \bibnamefont
  {Siemß}}, \bibinfo {author} {\bibfnamefont {F.}~\bibnamefont {Fitzek}},
  \bibinfo {author} {\bibfnamefont {S.}~\bibnamefont {Abend}}, \bibinfo
  {author} {\bibfnamefont {E.~M.}\ \bibnamefont {Rasel}}, \bibinfo {author}
  {\bibfnamefont {N.}~\bibnamefont {Gaaloul}},\ and\ \bibinfo {author}
  {\bibfnamefont {K.}~\bibnamefont {Hammerer}},\ }\bibfield  {title} {\bibinfo
  {title} {Analytic theory for {B}ragg atom interferometry based on the
  adiabatic theorem},\ }\href {https://doi.org/10.1103/PhysRevA.102.033709}
  {\bibfield  {journal} {\bibinfo  {journal} {Phys. Rev. A}\ }\textbf {\bibinfo
  {volume} {102}},\ \bibinfo {pages} {033709} (\bibinfo {year}
  {2020})}\BibitemShut {NoStop}%
\bibitem [{\citenamefont {Chiow}\ \emph {et~al.}(2011)\citenamefont {Chiow},
  \citenamefont {Kovachy}, \citenamefont {Chien},\ and\ \citenamefont
  {Kasevich}}]{chiow_102hbark_2011}%
  \BibitemOpen
  \bibfield  {author} {\bibinfo {author} {\bibfnamefont {S.-w.}\ \bibnamefont
  {Chiow}}, \bibinfo {author} {\bibfnamefont {T.}~\bibnamefont {Kovachy}},
  \bibinfo {author} {\bibfnamefont {H.-C.}\ \bibnamefont {Chien}},\ and\
  \bibinfo {author} {\bibfnamefont {M.~A.}\ \bibnamefont {Kasevich}},\
  }\bibfield  {title} {\bibinfo {title} {{$102 \hbar k$ Large Area Atom
  Interferometers}},\ }\href {https://doi.org/10.1103/PhysRevLett.107.130403}
  {\bibfield  {journal} {\bibinfo  {journal} {Phys. Rev. Lett.}\ }\textbf
  {\bibinfo {volume} {107}},\ \bibinfo {pages} {130403} (\bibinfo {year}
  {2011})}\BibitemShut {NoStop}%
\bibitem [{\citenamefont {Gebbe}\ \emph {et~al.}(2021)\citenamefont {Gebbe},
  \citenamefont {Siemß}, \citenamefont {Gersemann}, \citenamefont {Müntinga},
  \citenamefont {Herrmann}, \citenamefont {Lämmerzahl}, \citenamefont
  {Ahlers}, \citenamefont {Gaaloul}, \citenamefont {Schubert}, \citenamefont
  {Hammerer}, \citenamefont {Abend},\ and\ \citenamefont
  {Rasel}}]{gebbe_twin-lattice_2021}%
  \BibitemOpen
  \bibfield  {author} {\bibinfo {author} {\bibfnamefont {M.}~\bibnamefont
  {Gebbe}}, \bibinfo {author} {\bibfnamefont {J.-N.}\ \bibnamefont {Siemß}},
  \bibinfo {author} {\bibfnamefont {M.}~\bibnamefont {Gersemann}}, \bibinfo
  {author} {\bibfnamefont {H.}~\bibnamefont {Müntinga}}, \bibinfo {author}
  {\bibfnamefont {S.}~\bibnamefont {Herrmann}}, \bibinfo {author}
  {\bibfnamefont {C.}~\bibnamefont {Lämmerzahl}}, \bibinfo {author}
  {\bibfnamefont {H.}~\bibnamefont {Ahlers}}, \bibinfo {author} {\bibfnamefont
  {N.}~\bibnamefont {Gaaloul}}, \bibinfo {author} {\bibfnamefont
  {C.}~\bibnamefont {Schubert}}, \bibinfo {author} {\bibfnamefont
  {K.}~\bibnamefont {Hammerer}}, \bibinfo {author} {\bibfnamefont
  {S.}~\bibnamefont {Abend}},\ and\ \bibinfo {author} {\bibfnamefont {E.~M.}\
  \bibnamefont {Rasel}},\ }\bibfield  {title} {\bibinfo {title} {Twin-lattice
  atom interferometry},\ }\href {https://doi.org/10.1038/s41467-021-22823-8}
  {\bibfield  {journal} {\bibinfo  {journal} {Nat. Commun.}\ }\textbf {\bibinfo
  {volume} {12}},\ \bibinfo {pages} {2544} (\bibinfo {year}
  {2021})}\BibitemShut {NoStop}%
\bibitem [{\citenamefont {Szigeti}\ \emph {et~al.}(2012)\citenamefont
  {Szigeti}, \citenamefont {Debs}, \citenamefont {Hope}, \citenamefont
  {Robins},\ and\ \citenamefont {Close}}]{Szigeti_2012}%
  \BibitemOpen
  \bibfield  {author} {\bibinfo {author} {\bibfnamefont {S.~S.}\ \bibnamefont
  {Szigeti}}, \bibinfo {author} {\bibfnamefont {J.~E.}\ \bibnamefont {Debs}},
  \bibinfo {author} {\bibfnamefont {J.~J.}\ \bibnamefont {Hope}}, \bibinfo
  {author} {\bibfnamefont {N.~P.}\ \bibnamefont {Robins}},\ and\ \bibinfo
  {author} {\bibfnamefont {J.~D.}\ \bibnamefont {Close}},\ }\bibfield  {title}
  {\bibinfo {title} {Why momentum width matters for atom interferometry with
  {B}ragg pulses},\ }\href {https://doi.org/10.1088/1367-2630/14/2/023009}
  {\bibfield  {journal} {\bibinfo  {journal} {New J. Phys.}\ }\textbf {\bibinfo
  {volume} {14}},\ \bibinfo {pages} {023009} (\bibinfo {year}
  {2012})}\BibitemShut {NoStop}%
\bibitem [{\citenamefont {Hartmann}\ \emph
  {et~al.}(2020{\natexlab{a}})\citenamefont {Hartmann}, \citenamefont
  {Jenewein}, \citenamefont {Giese}, \citenamefont {Abend}, \citenamefont
  {Roura}, \citenamefont {Rasel},\ and\ \citenamefont
  {Schleich}}]{Hartmann2020a}%
  \BibitemOpen
  \bibfield  {author} {\bibinfo {author} {\bibfnamefont {S.}~\bibnamefont
  {Hartmann}}, \bibinfo {author} {\bibfnamefont {J.}~\bibnamefont {Jenewein}},
  \bibinfo {author} {\bibfnamefont {E.}~\bibnamefont {Giese}}, \bibinfo
  {author} {\bibfnamefont {S.}~\bibnamefont {Abend}}, \bibinfo {author}
  {\bibfnamefont {A.}~\bibnamefont {Roura}}, \bibinfo {author} {\bibfnamefont
  {E.~M.}\ \bibnamefont {Rasel}},\ and\ \bibinfo {author} {\bibfnamefont
  {W.~P.}\ \bibnamefont {Schleich}},\ }\bibfield  {title} {\bibinfo {title}
  {Regimes of atomic diffraction: {R}aman versus {B}ragg diffraction in
  retroreflective geometries},\ }\href
  {https://doi.org/10.1103/PhysRevA.101.053610} {\bibfield  {journal} {\bibinfo
   {journal} {Phys. Rev. A}\ }\textbf {\bibinfo {volume} {101}},\ \bibinfo
  {pages} {053610} (\bibinfo {year} {2020}{\natexlab{a}})}\BibitemShut
  {NoStop}%
\bibitem [{\citenamefont {Neumann}\ \emph {et~al.}(2021)\citenamefont
  {Neumann}, \citenamefont {Gebbe},\ and\ \citenamefont
  {Walser}}]{neumann_aberrations_2021}%
  \BibitemOpen
  \bibfield  {author} {\bibinfo {author} {\bibfnamefont {A.}~\bibnamefont
  {Neumann}}, \bibinfo {author} {\bibfnamefont {M.}~\bibnamefont {Gebbe}},\
  and\ \bibinfo {author} {\bibfnamefont {R.}~\bibnamefont {Walser}},\
  }\bibfield  {title} {\bibinfo {title} {Aberrations in (3+1)-dimensional
  {Bragg} diffraction using pulsed {Laguerre-Gaussian} laser beams},\ }\href
  {https://doi.org/10.1103/PhysRevA.103.043306} {\bibfield  {journal} {\bibinfo
   {journal} {Phys. Rev. A}\ }\textbf {\bibinfo {volume} {103}},\ \bibinfo
  {pages} {043306} (\bibinfo {year} {2021})}\BibitemShut {NoStop}%
\bibitem [{\citenamefont {Manna}(2020)}]{manna_nonadiabatic_2020}%
  \BibitemOpen
  \bibfield  {author} {\bibinfo {author} {\bibfnamefont {D.}~\bibnamefont
  {Manna}},\ }\bibfield  {title} {\bibinfo {title} {Nonadiabatic contributions
  to {{Bragg-regime}} dynamics in atomic {{Kapitza-Dirac}} scattering},\ }\href
  {https://doi.org/10.1103/PhysRevA.101.063621} {\bibfield  {journal} {\bibinfo
   {journal} {Phys. Rev. A}\ }\textbf {\bibinfo {volume} {101}},\ \bibinfo
  {pages} {063621} (\bibinfo {year} {2020})}\BibitemShut {NoStop}%
\bibitem [{\citenamefont {Lu}\ \emph {et~al.}(2018)\citenamefont {Lu},
  \citenamefont {Yao}, \citenamefont {Li}, \citenamefont {Luo}, \citenamefont
  {Barthwal}, \citenamefont {Chen}, \citenamefont {Lu}, \citenamefont {Wang},\
  and\ \citenamefont {Zhan}}]{lu_competition_2018}%
  \BibitemOpen
  \bibfield  {author} {\bibinfo {author} {\bibfnamefont {S.-B.}\ \bibnamefont
  {Lu}}, \bibinfo {author} {\bibfnamefont {Z.-W.}\ \bibnamefont {Yao}},
  \bibinfo {author} {\bibfnamefont {R.-B.}\ \bibnamefont {Li}}, \bibinfo
  {author} {\bibfnamefont {J.}~\bibnamefont {Luo}}, \bibinfo {author}
  {\bibfnamefont {S.}~\bibnamefont {Barthwal}}, \bibinfo {author}
  {\bibfnamefont {H.-H.}\ \bibnamefont {Chen}}, \bibinfo {author}
  {\bibfnamefont {Z.-X.}\ \bibnamefont {Lu}}, \bibinfo {author} {\bibfnamefont
  {J.}~\bibnamefont {Wang}},\ and\ \bibinfo {author} {\bibfnamefont {M.-S.}\
  \bibnamefont {Zhan}},\ }\bibfield  {title} {\bibinfo {title} {Competition
  effects of multiple quantum paths in an atom interferometer},\ }\href
  {https://doi.org/10.1016/j.optcom.2018.08.016} {\bibfield  {journal}
  {\bibinfo  {journal} {Opt. Commun.}\ }\textbf {\bibinfo {volume} {429}},\
  \bibinfo {pages} {158} (\bibinfo {year} {2018})}\BibitemShut {NoStop}%
\bibitem [{\citenamefont {Altin}\ \emph {et~al.}(2013)\citenamefont {Altin},
  \citenamefont {Johnsson}, \citenamefont {Negnevitsky}, \citenamefont
  {Dennis}, \citenamefont {Anderson}, \citenamefont {Debs}, \citenamefont
  {Szigeti}, \citenamefont {Hardman}, \citenamefont {Bennetts}, \citenamefont
  {McDonald}, \citenamefont {Turner}, \citenamefont {Close},\ and\
  \citenamefont {Robins}}]{altin_precision_2013}%
  \BibitemOpen
  \bibfield  {author} {\bibinfo {author} {\bibfnamefont {P.~A.}\ \bibnamefont
  {Altin}}, \bibinfo {author} {\bibfnamefont {M.~T.}\ \bibnamefont {Johnsson}},
  \bibinfo {author} {\bibfnamefont {V.}~\bibnamefont {Negnevitsky}}, \bibinfo
  {author} {\bibfnamefont {G.~R.}\ \bibnamefont {Dennis}}, \bibinfo {author}
  {\bibfnamefont {R.~P.}\ \bibnamefont {Anderson}}, \bibinfo {author}
  {\bibfnamefont {J.~E.}\ \bibnamefont {Debs}}, \bibinfo {author}
  {\bibfnamefont {S.~S.}\ \bibnamefont {Szigeti}}, \bibinfo {author}
  {\bibfnamefont {K.~S.}\ \bibnamefont {Hardman}}, \bibinfo {author}
  {\bibfnamefont {S.}~\bibnamefont {Bennetts}}, \bibinfo {author}
  {\bibfnamefont {G.~D.}\ \bibnamefont {McDonald}}, \bibinfo {author}
  {\bibfnamefont {L.~D.}\ \bibnamefont {Turner}}, \bibinfo {author}
  {\bibfnamefont {J.~D.}\ \bibnamefont {Close}},\ and\ \bibinfo {author}
  {\bibfnamefont {N.~P.}\ \bibnamefont {Robins}},\ }\bibfield  {title}
  {\bibinfo {title} {Precision atomic gravimeter based on {{Bragg}}
  diffraction},\ }\href {https://doi.org/10.1088/1367-2630/15/2/023009}
  {\bibfield  {journal} {\bibinfo  {journal} {New J. Phys.}\ }\textbf {\bibinfo
  {volume} {15}},\ \bibinfo {pages} {023009} (\bibinfo {year}
  {2013})}\BibitemShut {NoStop}%
\bibitem [{\citenamefont {Plotkin-Swing}\ \emph {et~al.}(2022)\citenamefont
  {Plotkin-Swing}, \citenamefont {Gochnauer}, \citenamefont {{McAlpine}},
  \citenamefont {Cooper}, \citenamefont {Jamison},\ and\ \citenamefont
  {Gupta}}]{plotkin-swing_three-path_2018}%
  \BibitemOpen
  \bibfield  {author} {\bibinfo {author} {\bibfnamefont {B.}~\bibnamefont
  {Plotkin-Swing}}, \bibinfo {author} {\bibfnamefont {D.}~\bibnamefont
  {Gochnauer}}, \bibinfo {author} {\bibfnamefont {K.~E.}\ \bibnamefont
  {{McAlpine}}}, \bibinfo {author} {\bibfnamefont {E.~S.}\ \bibnamefont
  {Cooper}}, \bibinfo {author} {\bibfnamefont {A.~O.}\ \bibnamefont
  {Jamison}},\ and\ \bibinfo {author} {\bibfnamefont {S.}~\bibnamefont
  {Gupta}},\ }\bibfield  {title} {\bibinfo {title} {{Three-Path Atom
  Interferometry with Large Momentum Separation}},\ }\href
  {https://doi.org/10.1103/PhysRevLett.121.133201} {\bibfield  {journal}
  {\bibinfo  {journal} {Phys. Rev. Lett.}\ }\textbf {\bibinfo {volume} {121}},\
  \bibinfo {pages} {133201} (\bibinfo {year} {2022})}\BibitemShut {NoStop}%
\bibitem [{\citenamefont {Parker}\ \emph {et~al.}(2016)\citenamefont {Parker},
  \citenamefont {Yu}, \citenamefont {Estey}, \citenamefont {Zhong},
  \citenamefont {Huang},\ and\ \citenamefont
  {Müller}}]{parker_controlling_2016}%
  \BibitemOpen
  \bibfield  {author} {\bibinfo {author} {\bibfnamefont {R.~H.}\ \bibnamefont
  {Parker}}, \bibinfo {author} {\bibfnamefont {C.}~\bibnamefont {Yu}}, \bibinfo
  {author} {\bibfnamefont {B.}~\bibnamefont {Estey}}, \bibinfo {author}
  {\bibfnamefont {W.}~\bibnamefont {Zhong}}, \bibinfo {author} {\bibfnamefont
  {E.}~\bibnamefont {Huang}},\ and\ \bibinfo {author} {\bibfnamefont
  {H.}~\bibnamefont {Müller}},\ }\bibfield  {title} {\bibinfo {title}
  {Controlling the multiport nature of {Bragg} diffraction in atom
  interferometry},\ }\href {https://doi.org/10.1103/PhysRevA.94.053618}
  {\bibfield  {journal} {\bibinfo  {journal} {Phys. Rev. A}\ }\textbf {\bibinfo
  {volume} {94}},\ \bibinfo {pages} {053618} (\bibinfo {year}
  {2016})}\BibitemShut {NoStop}%
\bibitem [{\citenamefont {He}\ \emph {et~al.}(2021)\citenamefont {He},
  \citenamefont {Chen}, \citenamefont {Fang}, \citenamefont {Ge}, \citenamefont
  {Li}, \citenamefont {Zhang}, \citenamefont {Zhou}, \citenamefont {Wang},\
  and\ \citenamefont {Zhan}}]{he_phase_2021}%
  \BibitemOpen
  \bibfield  {author} {\bibinfo {author} {\bibfnamefont {M.}~\bibnamefont
  {He}}, \bibinfo {author} {\bibfnamefont {X.}~\bibnamefont {Chen}}, \bibinfo
  {author} {\bibfnamefont {J.}~\bibnamefont {Fang}}, \bibinfo {author}
  {\bibfnamefont {G.}~\bibnamefont {Ge}}, \bibinfo {author} {\bibfnamefont
  {J.}~\bibnamefont {Li}}, \bibinfo {author} {\bibfnamefont {D.}~\bibnamefont
  {Zhang}}, \bibinfo {author} {\bibfnamefont {L.}~\bibnamefont {Zhou}},
  \bibinfo {author} {\bibfnamefont {J.}~\bibnamefont {Wang}},\ and\ \bibinfo
  {author} {\bibfnamefont {M.}~\bibnamefont {Zhan}},\ }\bibfield  {title}
  {\bibinfo {title} {Phase shift of double-diffraction {R}aman interference due
  to high-order diffraction states},\ }\href
  {https://link.aps.org/doi/10.1103/PhysRevA.103.063310} {\bibfield  {journal}
  {\bibinfo  {journal} {Phys. Rev. A}\ }\textbf {\bibinfo {volume} {103}},\
  \bibinfo {pages} {063310} (\bibinfo {year} {2021})}\BibitemShut {NoStop}%
\bibitem [{\citenamefont {Giltner}\ \emph {et~al.}(1995)\citenamefont
  {Giltner}, \citenamefont {{McGowan}},\ and\ \citenamefont
  {Lee}}]{giltner_theoretical_1995}%
  \BibitemOpen
  \bibfield  {author} {\bibinfo {author} {\bibfnamefont {D.~M.}\ \bibnamefont
  {Giltner}}, \bibinfo {author} {\bibfnamefont {R.~W.}\ \bibnamefont
  {{McGowan}}},\ and\ \bibinfo {author} {\bibfnamefont {S.~A.}\ \bibnamefont
  {Lee}},\ }\bibfield  {title} {\bibinfo {title} {Theoretical and experimental
  study of the {Bragg} scattering of atoms from a standing light wave},\ }\href
  {https://doi.org/10.1103/PhysRevA.52.3966} {\bibfield  {journal} {\bibinfo
  {journal} {Phys. Rev. A}\ }\textbf {\bibinfo {volume} {52}},\ \bibinfo
  {pages} {3966} (\bibinfo {year} {1995})}\BibitemShut {NoStop}%
\bibitem [{\citenamefont {D{\" u}rr}\ and\ \citenamefont
  {Rempe}(1999)}]{durr_acceptance_1999}%
  \BibitemOpen
  \bibfield  {author} {\bibinfo {author} {\bibfnamefont {S.}~\bibnamefont {D{\"
  u}rr}}\ and\ \bibinfo {author} {\bibfnamefont {G.}~\bibnamefont {Rempe}},\
  }\bibfield  {title} {\bibinfo {title} {Acceptance angle for {B}ragg
  reflection of atoms from a standing light wave},\ }\href
  {https://doi.org/10.1103/PhysRevA.59.1495} {\bibfield  {journal} {\bibinfo
  {journal} {Phys. Rev. A}\ }\textbf {\bibinfo {volume} {59}},\ \bibinfo
  {pages} {1495} (\bibinfo {year} {1999})}\BibitemShut {NoStop}%
\bibitem [{\citenamefont {Moler}\ \emph {et~al.}(1992)\citenamefont {Moler},
  \citenamefont {Weiss}, \citenamefont {Kasevich},\ and\ \citenamefont
  {Chu}}]{moler_theoretical_1992}%
  \BibitemOpen
  \bibfield  {author} {\bibinfo {author} {\bibfnamefont {K.}~\bibnamefont
  {Moler}}, \bibinfo {author} {\bibfnamefont {D.~S.}\ \bibnamefont {Weiss}},
  \bibinfo {author} {\bibfnamefont {M.~A.}\ \bibnamefont {Kasevich}},\ and\
  \bibinfo {author} {\bibfnamefont {S.}~\bibnamefont {Chu}},\ }\bibfield
  {title} {\bibinfo {title} {Theoretical analysis of velocity-selective {R}aman
  transitions},\ }\href {https://doi.org/10.1103/PhysRevA.45.342} {\bibfield
  {journal} {\bibinfo  {journal} {Phys. Rev. A}\ }\textbf {\bibinfo {volume}
  {45}},\ \bibinfo {pages} {342} (\bibinfo {year} {1992})}\BibitemShut
  {NoStop}%
\bibitem [{\citenamefont {Torii}\ \emph {et~al.}(2000)\citenamefont {Torii},
  \citenamefont {Suzuki}, \citenamefont {Kozuma}, \citenamefont {Sugiura},
  \citenamefont {Kuga}, \citenamefont {Deng},\ and\ \citenamefont
  {Hagley}}]{torii_mach-zehnder_2000}%
  \BibitemOpen
  \bibfield  {author} {\bibinfo {author} {\bibfnamefont {Y.}~\bibnamefont
  {Torii}}, \bibinfo {author} {\bibfnamefont {Y.}~\bibnamefont {Suzuki}},
  \bibinfo {author} {\bibfnamefont {M.}~\bibnamefont {Kozuma}}, \bibinfo
  {author} {\bibfnamefont {T.}~\bibnamefont {Sugiura}}, \bibinfo {author}
  {\bibfnamefont {T.}~\bibnamefont {Kuga}}, \bibinfo {author} {\bibfnamefont
  {L.}~\bibnamefont {Deng}},\ and\ \bibinfo {author} {\bibfnamefont {E.~W.}\
  \bibnamefont {Hagley}},\ }\bibfield  {title} {\bibinfo {title}
  {Mach-{Zehnder} {Bragg} interferometer for a {Bose}-{Einstein} condensate},\
  }\href {https://doi.org/10.1103/PhysRevA.61.041602} {\bibfield  {journal}
  {\bibinfo  {journal} {Phys. Rev. A}\ }\textbf {\bibinfo {volume} {61}},\
  \bibinfo {pages} {041602} (\bibinfo {year} {2000})}\BibitemShut {NoStop}%
\bibitem [{\citenamefont {Giese}(2015)}]{giese2015mechanisms}%
  \BibitemOpen
  \bibfield  {author} {\bibinfo {author} {\bibfnamefont {E.}~\bibnamefont
  {Giese}},\ }\bibfield  {title} {\bibinfo {title} {Mechanisms of matter-wave
  diffraction and their application to interferometers},\ }\href
  {https://onlinelibrary.wiley.com/doi/full/10.1002/prop.201500020} {\bibfield
  {journal} {\bibinfo  {journal} {Fortschr. Phys.}\ }\textbf {\bibinfo {volume}
  {63}},\ \bibinfo {pages} {337} (\bibinfo {year} {2015})}\BibitemShut
  {NoStop}%
\bibitem [{\citenamefont {Müller}\ \emph
  {et~al.}(2008{\natexlab{b}})\citenamefont {Müller}, \citenamefont {Chiow},\
  and\ \citenamefont {Chu}}]{muller_atom-wave_2008}%
  \BibitemOpen
  \bibfield  {author} {\bibinfo {author} {\bibfnamefont {H.}~\bibnamefont
  {Müller}}, \bibinfo {author} {\bibfnamefont {S.-w.}\ \bibnamefont {Chiow}},\
  and\ \bibinfo {author} {\bibfnamefont {S.}~\bibnamefont {Chu}},\ }\bibfield
  {title} {\bibinfo {title} {Atom-wave diffraction between the {{Raman-Nath}}
  and the {{Bragg}} regime: {{Effective Rabi}} frequency, losses, and phase
  shifts},\ }\href {https://doi.org/10.1103/PhysRevA.77.023609} {\bibfield
  {journal} {\bibinfo  {journal} {Phys. Rev. A}\ }\textbf {\bibinfo {volume}
  {77}},\ \bibinfo {pages} {023609} (\bibinfo {year}
  {2008}{\natexlab{b}})}\BibitemShut {NoStop}%
\bibitem [{\citenamefont {Gould}\ \emph {et~al.}(1986)\citenamefont {Gould},
  \citenamefont {Ruff},\ and\ \citenamefont
  {Pritchard}}]{gould1986diffraction}%
  \BibitemOpen
  \bibfield  {author} {\bibinfo {author} {\bibfnamefont {P.~L.}\ \bibnamefont
  {Gould}}, \bibinfo {author} {\bibfnamefont {G.~A.}\ \bibnamefont {Ruff}},\
  and\ \bibinfo {author} {\bibfnamefont {D.~E.}\ \bibnamefont {Pritchard}},\
  }\bibfield  {title} {\bibinfo {title} {Diffraction of atoms by light: {T}he
  near-resonant {K}apitza-{D}irac effect},\ }\href
  {https://doi.org/10.1103/PhysRevLett.56.827} {\bibfield  {journal} {\bibinfo
  {journal} {Phys. Rev. Lett.}\ }\textbf {\bibinfo {volume} {56}},\ \bibinfo
  {pages} {827} (\bibinfo {year} {1986})}\BibitemShut {NoStop}%
\bibitem [{\citenamefont {Béguin}\ \emph {et~al.}(2022)\citenamefont
  {Béguin}, \citenamefont {Rodzinka}, \citenamefont {Vigué}, \citenamefont
  {Allard},\ and\ \citenamefont {Gauguet}}]{beguin_characterization_2021}%
  \BibitemOpen
  \bibfield  {author} {\bibinfo {author} {\bibfnamefont {A.}~\bibnamefont
  {Béguin}}, \bibinfo {author} {\bibfnamefont {T.}~\bibnamefont {Rodzinka}},
  \bibinfo {author} {\bibfnamefont {J.}~\bibnamefont {Vigué}}, \bibinfo
  {author} {\bibfnamefont {B.}~\bibnamefont {Allard}},\ and\ \bibinfo {author}
  {\bibfnamefont {A.}~\bibnamefont {Gauguet}},\ }\bibfield  {title} {\bibinfo
  {title} {Characterization of an atom interferometer in the {quasi-Bragg}
  regime},\ }\href {https://doi.org/10.1103/PhysRevA.105.033302} {\bibfield
  {journal} {\bibinfo  {journal} {Phys. Rev. A}\ }\textbf {\bibinfo {volume}
  {105}},\ \bibinfo {pages} {033302} (\bibinfo {year} {2022})}\BibitemShut
  {NoStop}%
\bibitem [{\citenamefont {Carraz}\ \emph {et~al.}(2012)\citenamefont {Carraz},
  \citenamefont {Charrière}, \citenamefont {Cadoret}, \citenamefont {Zahzam},
  \citenamefont {Bidel},\ and\ \citenamefont {Bresson}}]{carraz_phase_2012}%
  \BibitemOpen
  \bibfield  {author} {\bibinfo {author} {\bibfnamefont {O.}~\bibnamefont
  {Carraz}}, \bibinfo {author} {\bibfnamefont {R.}~\bibnamefont {Charrière}},
  \bibinfo {author} {\bibfnamefont {M.}~\bibnamefont {Cadoret}}, \bibinfo
  {author} {\bibfnamefont {N.}~\bibnamefont {Zahzam}}, \bibinfo {author}
  {\bibfnamefont {Y.}~\bibnamefont {Bidel}},\ and\ \bibinfo {author}
  {\bibfnamefont {A.}~\bibnamefont {Bresson}},\ }\bibfield  {title} {\bibinfo
  {title} {Phase shift in an atom interferometer induced by the additional
  laser lines of a {Raman} laser generated by modulation},\ }\href
  {https://doi.org/10.1103/PhysRevA.86.033605} {\bibfield  {journal} {\bibinfo
  {journal} {Phys. Rev. A}\ }\textbf {\bibinfo {volume} {86}},\ \bibinfo
  {pages} {033605} (\bibinfo {year} {2012})}\BibitemShut {NoStop}%
\bibitem [{\citenamefont {Gauguet}\ \emph {et~al.}(2008)\citenamefont
  {Gauguet}, \citenamefont {Mehlstäubler}, \citenamefont {Lévèque},
  \citenamefont {Le~Gouët}, \citenamefont {Chaibi}, \citenamefont {Canuel},
  \citenamefont {Clairon}, \citenamefont {Pereira Dos~Santos},\ and\
  \citenamefont {Landragin}}]{gauguet_off-resonant_2008}%
  \BibitemOpen
  \bibfield  {author} {\bibinfo {author} {\bibfnamefont {A.}~\bibnamefont
  {Gauguet}}, \bibinfo {author} {\bibfnamefont {T.~E.}\ \bibnamefont
  {Mehlstäubler}}, \bibinfo {author} {\bibfnamefont {T.}~\bibnamefont
  {Lévèque}}, \bibinfo {author} {\bibfnamefont {J.}~\bibnamefont
  {Le~Gouët}}, \bibinfo {author} {\bibfnamefont {W.}~\bibnamefont {Chaibi}},
  \bibinfo {author} {\bibfnamefont {B.}~\bibnamefont {Canuel}}, \bibinfo
  {author} {\bibfnamefont {A.}~\bibnamefont {Clairon}}, \bibinfo {author}
  {\bibfnamefont {F.}~\bibnamefont {Pereira Dos~Santos}},\ and\ \bibinfo
  {author} {\bibfnamefont {A.}~\bibnamefont {Landragin}},\ }\bibfield  {title}
  {\bibinfo {title} {Off-resonant {Raman} transition impact in an atom
  interferometer},\ }\href {https://doi.org/10.1103/PhysRevA.78.043615}
  {\bibfield  {journal} {\bibinfo  {journal} {Phys. Rev. A}\ }\textbf {\bibinfo
  {volume} {78}},\ \bibinfo {pages} {043615} (\bibinfo {year}
  {2008})}\BibitemShut {NoStop}%
\bibitem [{\citenamefont {Giese}\ \emph {et~al.}(2016)\citenamefont {Giese},
  \citenamefont {Friedrich}, \citenamefont {Abend}, \citenamefont {Rasel},\
  and\ \citenamefont {Schleich}}]{giese_light_2016}%
  \BibitemOpen
  \bibfield  {author} {\bibinfo {author} {\bibfnamefont {E.}~\bibnamefont
  {Giese}}, \bibinfo {author} {\bibfnamefont {A.}~\bibnamefont {Friedrich}},
  \bibinfo {author} {\bibfnamefont {S.}~\bibnamefont {Abend}}, \bibinfo
  {author} {\bibfnamefont {E.~M.}\ \bibnamefont {Rasel}},\ and\ \bibinfo
  {author} {\bibfnamefont {W.~P.}\ \bibnamefont {Schleich}},\ }\bibfield
  {title} {\bibinfo {title} {Light shifts in atomic {{Bragg}} diffraction},\
  }\href {https://doi.org/10.1103/PhysRevA.94.063619} {\bibfield  {journal}
  {\bibinfo  {journal} {Phys. Rev. A}\ }\textbf {\bibinfo {volume} {94}},\
  \bibinfo {pages} {063619} (\bibinfo {year} {2016})}\BibitemShut {NoStop}%
\bibitem [{\citenamefont {Gochnauer}\ \emph {et~al.}(2019)\citenamefont
  {Gochnauer}, \citenamefont {{McAlpine}}, \citenamefont {Plotkin-Swing},
  \citenamefont {Jamison},\ and\ \citenamefont
  {Gupta}}]{gochnauer_bloch-band_2019}%
  \BibitemOpen
  \bibfield  {author} {\bibinfo {author} {\bibfnamefont {D.}~\bibnamefont
  {Gochnauer}}, \bibinfo {author} {\bibfnamefont {K.~E.}\ \bibnamefont
  {{McAlpine}}}, \bibinfo {author} {\bibfnamefont {B.}~\bibnamefont
  {Plotkin-Swing}}, \bibinfo {author} {\bibfnamefont {A.~O.}\ \bibnamefont
  {Jamison}},\ and\ \bibinfo {author} {\bibfnamefont {S.}~\bibnamefont
  {Gupta}},\ }\bibfield  {title} {\bibinfo {title} {Bloch-band picture for
  light-pulse atom diffraction and interferometry},\ }\href
  {https://doi.org/10.1103/PhysRevA.100.043611} {\bibfield  {journal} {\bibinfo
   {journal} {Phys. Rev. A}\ }\textbf {\bibinfo {volume} {100}},\ \bibinfo
  {pages} {043611} (\bibinfo {year} {2019})}\BibitemShut {NoStop}%
\bibitem [{\citenamefont {Estey}\ \emph {et~al.}(2015)\citenamefont {Estey},
  \citenamefont {Yu}, \citenamefont {Müller}, \citenamefont {Kuan},\ and\
  \citenamefont {Lan}}]{estey_high-resolution_2015}%
  \BibitemOpen
  \bibfield  {author} {\bibinfo {author} {\bibfnamefont {B.}~\bibnamefont
  {Estey}}, \bibinfo {author} {\bibfnamefont {C.}~\bibnamefont {Yu}}, \bibinfo
  {author} {\bibfnamefont {H.}~\bibnamefont {Müller}}, \bibinfo {author}
  {\bibfnamefont {P.-C.}\ \bibnamefont {Kuan}},\ and\ \bibinfo {author}
  {\bibfnamefont {S.-Y.}\ \bibnamefont {Lan}},\ }\bibfield  {title} {\bibinfo
  {title} {{High-Resolution Atom Interferometers with Suppressed Diffraction
  Phases}},\ }\href {https://doi.org/10.1103/PhysRevLett.115.083002} {\bibfield
   {journal} {\bibinfo  {journal} {Phys. Rev. Lett.}\ }\textbf {\bibinfo
  {volume} {115}},\ \bibinfo {pages} {083002} (\bibinfo {year}
  {2015})}\BibitemShut {NoStop}%
\bibitem [{\citenamefont {Li}\ \emph {et~al.}(2015)\citenamefont {Li},
  \citenamefont {Shao},\ and\ \citenamefont {Hu}}]{li_raman_2015}%
  \BibitemOpen
  \bibfield  {author} {\bibinfo {author} {\bibfnamefont {X.}~\bibnamefont
  {Li}}, \bibinfo {author} {\bibfnamefont {C.-G.}\ \bibnamefont {Shao}},\ and\
  \bibinfo {author} {\bibfnamefont {Z.-K.}\ \bibnamefont {Hu}},\ }\bibfield
  {title} {\bibinfo {title} {Raman pulse duration effect in high-precision atom
  interferometry gravimeters},\ }\href
  {https://doi.org/10.1364/JOSAB.32.000248} {\bibfield  {journal} {\bibinfo
  {journal} {J. Opt. Soc. Am. B, JOSAB}\ }\textbf {\bibinfo {volume} {32}},\
  \bibinfo {pages} {248} (\bibinfo {year} {2015})}\BibitemShut {NoStop}%
\bibitem [{\citenamefont {Bertoldi}\ \emph {et~al.}(2019)\citenamefont
  {Bertoldi}, \citenamefont {Minardi},\ and\ \citenamefont
  {Prevedelli}}]{bertoldi_phase_2019}%
  \BibitemOpen
  \bibfield  {author} {\bibinfo {author} {\bibfnamefont {A.}~\bibnamefont
  {Bertoldi}}, \bibinfo {author} {\bibfnamefont {F.}~\bibnamefont {Minardi}},\
  and\ \bibinfo {author} {\bibfnamefont {M.}~\bibnamefont {Prevedelli}},\
  }\bibfield  {title} {\bibinfo {title} {Phase shift in atom interferometers:
  {C}orrections for nonquadratic potentials and finite-duration laser pulses},\
  }\href {https://doi.org/10.1103/PhysRevA.99.033619} {\bibfield  {journal}
  {\bibinfo  {journal} {Phys. Rev. A}\ }\textbf {\bibinfo {volume} {99}},\
  \bibinfo {pages} {033619} (\bibinfo {year} {2019})}\BibitemShut {NoStop}%
\bibitem [{\citenamefont {Antoine}(2006)}]{antoine_matter_2006}%
  \BibitemOpen
  \bibfield  {author} {\bibinfo {author} {\bibfnamefont {C.}~\bibnamefont
  {Antoine}},\ }\bibfield  {title} {\bibinfo {title} {Matter wave beam
  splitters in gravito-inertial and trapping potentials: generalized ttt scheme
  for atom interferometry},\ }\href {https://doi.org/10.1007/s00340-006-2378-8}
  {\bibfield  {journal} {\bibinfo  {journal} {Appl. Phys. B}\ }\textbf
  {\bibinfo {volume} {84}},\ \bibinfo {pages} {585} (\bibinfo {year}
  {2006})}\BibitemShut {NoStop}%
\bibitem [{\citenamefont {Peters}\ \emph {et~al.}(2001)\citenamefont {Peters},
  \citenamefont {Chung},\ and\ \citenamefont
  {Chu}}]{peters_high-precision_2001}%
  \BibitemOpen
  \bibfield  {author} {\bibinfo {author} {\bibfnamefont {A.}~\bibnamefont
  {Peters}}, \bibinfo {author} {\bibfnamefont {K.~Y.}\ \bibnamefont {Chung}},\
  and\ \bibinfo {author} {\bibfnamefont {S.}~\bibnamefont {Chu}},\ }\bibfield
  {title} {\bibinfo {title} {High-precision gravity measurements using atom
  interferometry},\ }\href {https://doi.org/10.1088/0026-1394/38/1/4}
  {\bibfield  {journal} {\bibinfo  {journal} {Metrologia}\ }\textbf {\bibinfo
  {volume} {38}},\ \bibinfo {pages} {25} (\bibinfo {year} {2001})}\BibitemShut
  {NoStop}%
\bibitem [{\citenamefont {Fitzek}\ \emph {et~al.}(2020)\citenamefont {Fitzek},
  \citenamefont {Siemß}, \citenamefont {Seckmeyer}, \citenamefont {Ahlers},
  \citenamefont {Rasel}, \citenamefont {Hammerer},\ and\ \citenamefont
  {Gaaloul}}]{fitzek_universal_2020}%
  \BibitemOpen
  \bibfield  {author} {\bibinfo {author} {\bibfnamefont {F.}~\bibnamefont
  {Fitzek}}, \bibinfo {author} {\bibfnamefont {J.-N.}\ \bibnamefont {Siemß}},
  \bibinfo {author} {\bibfnamefont {S.}~\bibnamefont {Seckmeyer}}, \bibinfo
  {author} {\bibfnamefont {H.}~\bibnamefont {Ahlers}}, \bibinfo {author}
  {\bibfnamefont {E.~M.}\ \bibnamefont {Rasel}}, \bibinfo {author}
  {\bibfnamefont {K.}~\bibnamefont {Hammerer}},\ and\ \bibinfo {author}
  {\bibfnamefont {N.}~\bibnamefont {Gaaloul}},\ }\bibfield  {title} {\bibinfo
  {title} {Universal atom interferometer simulation of elastic scattering
  processes},\ }\href {https://doi.org/10.1038/s41598-020-78859-1} {\bibfield
  {journal} {\bibinfo  {journal} {Sci. Rep.}\ }\textbf {\bibinfo {volume}
  {10}},\ \bibinfo {pages} {22120} (\bibinfo {year} {2020})}\BibitemShut
  {NoStop}%
\bibitem [{\citenamefont {Gupta}\ \emph {et~al.}(2002)\citenamefont {Gupta},
  \citenamefont {Dieckmann}, \citenamefont {Hadzibabic},\ and\ \citenamefont
  {Pritchard}}]{gupta_contrast_2002}%
  \BibitemOpen
  \bibfield  {author} {\bibinfo {author} {\bibfnamefont {S.}~\bibnamefont
  {Gupta}}, \bibinfo {author} {\bibfnamefont {K.}~\bibnamefont {Dieckmann}},
  \bibinfo {author} {\bibfnamefont {Z.}~\bibnamefont {Hadzibabic}},\ and\
  \bibinfo {author} {\bibfnamefont {D.~E.}\ \bibnamefont {Pritchard}},\
  }\bibfield  {title} {\bibinfo {title} {{Contrast Interferometry using
  Bose-Einstein Condensates to Measure $h/m$ and $\alpha$}},\ }\href
  {https://doi.org/10.1103/PhysRevLett.89.140401} {\bibfield  {journal}
  {\bibinfo  {journal} {Phys. Rev. Lett.}\ }\textbf {\bibinfo {volume} {89}},\
  \bibinfo {pages} {140401} (\bibinfo {year} {2002})}\BibitemShut {NoStop}%
\bibitem [{\citenamefont {He}\ \emph {et~al.}(2022)\citenamefont {He},
  \citenamefont {Ma},\ and\ \citenamefont {Li}}]{he_measuring_2022}%
  \BibitemOpen
  \bibfield  {author} {\bibinfo {author} {\bibfnamefont {T.-C.}\ \bibnamefont
  {He}}, \bibinfo {author} {\bibfnamefont {Y.-Q.}\ \bibnamefont {Ma}},\ and\
  \bibinfo {author} {\bibfnamefont {J.}~\bibnamefont {Li}},\ }\bibfield
  {title} {\bibinfo {title} {Measuring gravitational acceleration by cold atom
  multimode interference with three {{Kapitza}}–{{Dirac}} pulses},\ }\href
  {https://doi.org/10.1140/epjd/s10053-021-00335-w} {\bibfield  {journal}
  {\bibinfo  {journal} {Eur. Phys. J. D}\ }\textbf {\bibinfo {volume} {76}},\
  \bibinfo {pages} {3} (\bibinfo {year} {2022})}\BibitemShut {NoStop}%
\bibitem [{\citenamefont {Malossi}\ \emph {et~al.}(2010)\citenamefont
  {Malossi}, \citenamefont {Bodart}, \citenamefont {Merlet}, \citenamefont
  {L\'ev\`eque}, \citenamefont {Landragin},\ and\ \citenamefont {Pereira
  Dos~Santos}}]{Malossi_Double_2010}%
  \BibitemOpen
  \bibfield  {author} {\bibinfo {author} {\bibfnamefont {N.}~\bibnamefont
  {Malossi}}, \bibinfo {author} {\bibfnamefont {Q.}~\bibnamefont {Bodart}},
  \bibinfo {author} {\bibfnamefont {S.}~\bibnamefont {Merlet}}, \bibinfo
  {author} {\bibfnamefont {T.}~\bibnamefont {L\'ev\`eque}}, \bibinfo {author}
  {\bibfnamefont {A.}~\bibnamefont {Landragin}},\ and\ \bibinfo {author}
  {\bibfnamefont {F.}~\bibnamefont {Pereira Dos~Santos}},\ }\bibfield  {title}
  {\bibinfo {title} {Double diffraction in an atomic gravimeter},\ }\href
  {https://doi.org/10.1103/PhysRevA.81.013617} {\bibfield  {journal} {\bibinfo
  {journal} {Phys. Rev. A}\ }\textbf {\bibinfo {volume} {81}},\ \bibinfo
  {pages} {013617} (\bibinfo {year} {2010})}\BibitemShut {NoStop}%
\bibitem [{\citenamefont {Zhou}\ \emph {et~al.}(2015)\citenamefont {Zhou},
  \citenamefont {Long}, \citenamefont {Tang}, \citenamefont {Chen},
  \citenamefont {Gao}, \citenamefont {Peng}, \citenamefont {Duan},
  \citenamefont {Zhong}, \citenamefont {Xiong}, \citenamefont {Wang},
  \citenamefont {Zhang},\ and\ \citenamefont {Zhan}}]{zhou_test_2015}%
  \BibitemOpen
  \bibfield  {author} {\bibinfo {author} {\bibfnamefont {L.}~\bibnamefont
  {Zhou}}, \bibinfo {author} {\bibfnamefont {S.}~\bibnamefont {Long}}, \bibinfo
  {author} {\bibfnamefont {B.}~\bibnamefont {Tang}}, \bibinfo {author}
  {\bibfnamefont {X.}~\bibnamefont {Chen}}, \bibinfo {author} {\bibfnamefont
  {F.}~\bibnamefont {Gao}}, \bibinfo {author} {\bibfnamefont {W.}~\bibnamefont
  {Peng}}, \bibinfo {author} {\bibfnamefont {W.}~\bibnamefont {Duan}}, \bibinfo
  {author} {\bibfnamefont {J.}~\bibnamefont {Zhong}}, \bibinfo {author}
  {\bibfnamefont {Z.}~\bibnamefont {Xiong}}, \bibinfo {author} {\bibfnamefont
  {J.}~\bibnamefont {Wang}}, \bibinfo {author} {\bibfnamefont {Y.}~\bibnamefont
  {Zhang}},\ and\ \bibinfo {author} {\bibfnamefont {M.}~\bibnamefont {Zhan}},\
  }\bibfield  {title} {\bibinfo {title} {Test of {E}quivalence {P}rinciple at
  $1{0}^{\ensuremath{-}8}$ {L}evel by a {D}ual-{S}pecies {D}ouble-{D}iffraction
  {R}aman {A}tom {I}nterferometer},\ }\href
  {https://doi.org/10.1103/PhysRevLett.115.013004} {\bibfield  {journal}
  {\bibinfo  {journal} {Phys. Rev. Lett.}\ }\textbf {\bibinfo {volume} {115}},\
  \bibinfo {pages} {013004} (\bibinfo {year} {2015})}\BibitemShut {NoStop}%
\bibitem [{\citenamefont {Radmore}\ and\ \citenamefont
  {Knight}(1982)}]{radmore_population_1982}%
  \BibitemOpen
  \bibfield  {author} {\bibinfo {author} {\bibfnamefont {P.~M.}\ \bibnamefont
  {Radmore}}\ and\ \bibinfo {author} {\bibfnamefont {P.~L.}\ \bibnamefont
  {Knight}},\ }\bibfield  {title} {\bibinfo {title} {Population trapping and
  dispersion in a three-level system},\ }\href
  {https://doi.org/10.1088/0022-3700/15/4/009} {\bibfield  {journal} {\bibinfo
  {journal} {J. Phys. B: Atom. Mol. Phys.}\ }\textbf {\bibinfo {volume} {15}},\
  \bibinfo {pages} {561} (\bibinfo {year} {1982})}\BibitemShut {NoStop}%
\bibitem [{\citenamefont {McGuirk}\ \emph {et~al.}(2002)\citenamefont
  {McGuirk}, \citenamefont {Foster}, \citenamefont {Fixler}, \citenamefont
  {Snadden},\ and\ \citenamefont {Kasevich}}]{mcguirk_sensitive_2002}%
  \BibitemOpen
  \bibfield  {author} {\bibinfo {author} {\bibfnamefont {J.~M.}\ \bibnamefont
  {McGuirk}}, \bibinfo {author} {\bibfnamefont {G.~T.}\ \bibnamefont {Foster}},
  \bibinfo {author} {\bibfnamefont {J.~B.}\ \bibnamefont {Fixler}}, \bibinfo
  {author} {\bibfnamefont {M.~J.}\ \bibnamefont {Snadden}},\ and\ \bibinfo
  {author} {\bibfnamefont {M.~A.}\ \bibnamefont {Kasevich}},\ }\bibfield
  {title} {\bibinfo {title} {Sensitive absolute-gravity gradiometry using atom
  interferometry},\ }\href {https://doi.org/10.1103/PhysRevA.65.033608}
  {\bibfield  {journal} {\bibinfo  {journal} {Phys. Rev. A}\ }\textbf {\bibinfo
  {volume} {65}},\ \bibinfo {pages} {033608} (\bibinfo {year}
  {2002})}\BibitemShut {NoStop}%
\bibitem [{\citenamefont {Roura}(2020)}]{Roura2020}%
  \BibitemOpen
  \bibfield  {author} {\bibinfo {author} {\bibfnamefont {A.}~\bibnamefont
  {Roura}},\ }\bibfield  {title} {\bibinfo {title} {Gravitational redshift in
  quantum-clock interferometry},\ }\href
  {https://doi.org/10.1103/PhysRevX.10.021014} {\bibfield  {journal} {\bibinfo
  {journal} {Phys. Rev. X}\ }\textbf {\bibinfo {volume} {10}},\ \bibinfo
  {pages} {021014} (\bibinfo {year} {2020})}\BibitemShut {NoStop}%
\bibitem [{Note1()}]{Note1}%
  \BibitemOpen
  \bibinfo {note} {In both cases it is possible to have SBD rather than DBD,
  even for vanishing initial velocities, by simultaneously chirping both
  injected frequencies at a sufficiently high rate and exploiting the extra
  time of flight for the retroreflected components \cite {Perrin2019}.
  Alternatively, it is also possible to avoid DBD by considering magnetically
  sensitive states and suitable laser polarizations \cite
  {Bernard2022}.}\BibitemShut {Stop}%
\bibitem [{\citenamefont {Bord\'e}\ and\ \citenamefont
  {L\"ammerzahl}(1995)}]{lammerzahl_rabi_1995}%
  \BibitemOpen
  \bibfield  {author} {\bibinfo {author} {\bibfnamefont {C.~J.}\ \bibnamefont
  {Bord\'e}}\ and\ \bibinfo {author} {\bibfnamefont {C.}~\bibnamefont
  {L\"ammerzahl}},\ }\bibfield  {title} {\bibinfo {title} {Rabi oscillations in
  gravitational fields: {E}xact solution},\ }\href
  {https://doi.org/10.1016/0375-9601(95)00402-O} {\bibfield  {journal}
  {\bibinfo  {journal} {Phys. Lett. A}\ }\textbf {\bibinfo {volume} {203}},\
  \bibinfo {pages} {59} (\bibinfo {year} {1995})}\BibitemShut {NoStop}%
\bibitem [{\citenamefont {Marzlin}\ and\ \citenamefont
  {Audretsch}(1996)}]{marzlin_freely_1996}%
  \BibitemOpen
  \bibfield  {author} {\bibinfo {author} {\bibfnamefont {K.-P.}\ \bibnamefont
  {Marzlin}}\ and\ \bibinfo {author} {\bibfnamefont {J.}~\bibnamefont
  {Audretsch}},\ }\bibfield  {title} {\bibinfo {title} {{``Freely''} falling
  two-level atom in a running laser wave},\ }\href
  {https://doi.org/10.1103/PhysRevA.53.1004} {\bibfield  {journal} {\bibinfo
  {journal} {Phys. Rev. A}\ }\textbf {\bibinfo {volume} {53}},\ \bibinfo
  {pages} {1004} (\bibinfo {year} {1996})}\BibitemShut {NoStop}%
\bibitem [{\citenamefont {Roura}\ \emph {et~al.}(2014)\citenamefont {Roura},
  \citenamefont {Zeller},\ and\ \citenamefont {Schleich}}]{Roura2014}%
  \BibitemOpen
  \bibfield  {author} {\bibinfo {author} {\bibfnamefont {A.}~\bibnamefont
  {Roura}}, \bibinfo {author} {\bibfnamefont {W.}~\bibnamefont {Zeller}},\ and\
  \bibinfo {author} {\bibfnamefont {W.~P.}\ \bibnamefont {Schleich}},\
  }\bibfield  {title} {\bibinfo {title} {Overcoming loss of contrast in atom
  interferometry due to gravity gradients},\ }\href
  {http://stacks.iop.org/1367-2630/16/i=12/a=123012} {\bibfield  {journal}
  {\bibinfo  {journal} {New J. Phys.}\ }\textbf {\bibinfo {volume} {16}},\
  \bibinfo {pages} {123012} (\bibinfo {year} {2014})}\BibitemShut {NoStop}%
\bibitem [{\citenamefont {Hogan}\ \emph {et~al.}(2009)\citenamefont {Hogan},
  \citenamefont {Johnson},\ and\ \citenamefont {Kasevich}}]{Hogan2009}%
  \BibitemOpen
  \bibfield  {author} {\bibinfo {author} {\bibfnamefont {J.~M.}\ \bibnamefont
  {Hogan}}, \bibinfo {author} {\bibfnamefont {D.~M.~S.}\ \bibnamefont
  {Johnson}},\ and\ \bibinfo {author} {\bibfnamefont {M.~A.}\ \bibnamefont
  {Kasevich}},\ }\bibfield  {title} {\bibinfo {title} {Light-pulse atom
  interferometry},\ }in\ \href {https://doi.org/10.3254/978-1-58603-990-5-411}
  {\emph {\bibinfo {booktitle} {Atom Optics and Space Physics}}},\ \bibinfo
  {series} {Proceedings of the International School of Physics ``Enrico
  Fermi''}, Vol.\ \bibinfo {volume} {168},\ \bibinfo {editor} {edited by\
  \bibinfo {editor} {\bibfnamefont {E.}~\bibnamefont {Arimondo}}, \bibinfo
  {editor} {\bibfnamefont {W.}~\bibnamefont {Ertmer}}, \bibinfo {editor}
  {\bibfnamefont {E.~M.}\ \bibnamefont {Rasel}},\ and\ \bibinfo {editor}
  {\bibfnamefont {W.~P.}\ \bibnamefont {Schleich}}}\ (\bibinfo  {publisher}
  {IOS Press},\ \bibinfo {address} {Amsterdam},\ \bibinfo {year} {2009})\ pp.\
  \bibinfo {pages} {411 -- 447}\BibitemShut {NoStop}%
\bibitem [{\citenamefont {Lan}\ \emph {et~al.}(2012)\citenamefont {Lan},
  \citenamefont {Kuan}, \citenamefont {Estey}, \citenamefont {Haslinger},\ and\
  \citenamefont {M\"{u}ller}}]{Lan2012}%
  \BibitemOpen
  \bibfield  {author} {\bibinfo {author} {\bibfnamefont {S.-Y.}\ \bibnamefont
  {Lan}}, \bibinfo {author} {\bibfnamefont {P.-C.}\ \bibnamefont {Kuan}},
  \bibinfo {author} {\bibfnamefont {B.}~\bibnamefont {Estey}}, \bibinfo
  {author} {\bibfnamefont {P.}~\bibnamefont {Haslinger}},\ and\ \bibinfo
  {author} {\bibfnamefont {H.}~\bibnamefont {M\"{u}ller}},\ }\bibfield  {title}
  {\bibinfo {title} {Influence of the {C}oriolis force in atom
  interferometry},\ }\href {https://doi.org/10.1103/PhysRevLett.108.090402}
  {\bibfield  {journal} {\bibinfo  {journal} {Phys. Rev. Lett.}\ }\textbf
  {\bibinfo {volume} {108}},\ \bibinfo {pages} {090402} (\bibinfo {year}
  {2012})}\BibitemShut {NoStop}%
\bibitem [{\citenamefont {Dickerson}\ \emph {et~al.}(2013)\citenamefont
  {Dickerson}, \citenamefont {Hogan}, \citenamefont {Sugarbaker}, \citenamefont
  {Johnson},\ and\ \citenamefont {Kasevich}}]{Dickerson2013}%
  \BibitemOpen
  \bibfield  {author} {\bibinfo {author} {\bibfnamefont {S.~M.}\ \bibnamefont
  {Dickerson}}, \bibinfo {author} {\bibfnamefont {J.~M.}\ \bibnamefont
  {Hogan}}, \bibinfo {author} {\bibfnamefont {A.}~\bibnamefont {Sugarbaker}},
  \bibinfo {author} {\bibfnamefont {D.~M.~S.}\ \bibnamefont {Johnson}},\ and\
  \bibinfo {author} {\bibfnamefont {M.~A.}\ \bibnamefont {Kasevich}},\
  }\bibfield  {title} {\bibinfo {title} {Multiaxis inertial sensing with
  long-time point source atom interferometry},\ }\href
  {https://doi.org/10.1103/PhysRevLett.111.083001} {\bibfield  {journal}
  {\bibinfo  {journal} {Phys. Rev. Lett.}\ }\textbf {\bibinfo {volume} {111}},\
  \bibinfo {pages} {083001} (\bibinfo {year} {2013})}\BibitemShut {NoStop}%
\bibitem [{\citenamefont {Roura}(2017)}]{Roura2017}%
  \BibitemOpen
  \bibfield  {author} {\bibinfo {author} {\bibfnamefont {A.}~\bibnamefont
  {Roura}},\ }\bibfield  {title} {\bibinfo {title} {Circumventing
  {Heisenberg's} uncertainty principle in atom interferometry tests of the
  equivalence principle},\ }\href
  {https://doi.org/10.1103/PhysRevLett.118.160401} {\bibfield  {journal}
  {\bibinfo  {journal} {Phys. Rev. Lett.}\ }\textbf {\bibinfo {volume} {118}},\
  \bibinfo {pages} {160401} (\bibinfo {year} {2017})}\BibitemShut {NoStop}%
\bibitem [{\citenamefont {Overstreet}\ \emph {et~al.}(2018)\citenamefont
  {Overstreet}, \citenamefont {Asenbaum}, \citenamefont {Kovachy},
  \citenamefont {Notermans}, \citenamefont {Hogan},\ and\ \citenamefont
  {Kasevich}}]{Overstreet2018}%
  \BibitemOpen
  \bibfield  {author} {\bibinfo {author} {\bibfnamefont {C.}~\bibnamefont
  {Overstreet}}, \bibinfo {author} {\bibfnamefont {P.}~\bibnamefont
  {Asenbaum}}, \bibinfo {author} {\bibfnamefont {T.}~\bibnamefont {Kovachy}},
  \bibinfo {author} {\bibfnamefont {R.}~\bibnamefont {Notermans}}, \bibinfo
  {author} {\bibfnamefont {J.~M.}\ \bibnamefont {Hogan}},\ and\ \bibinfo
  {author} {\bibfnamefont {M.~A.}\ \bibnamefont {Kasevich}},\ }\bibfield
  {title} {\bibinfo {title} {Effective inertial frame in an atom
  interferometric test of the equivalence principle},\ }\href
  {https://doi.org/10.1103/PhysRevLett.120.183604} {\bibfield  {journal}
  {\bibinfo  {journal} {Phys. Rev. Lett.}\ }\textbf {\bibinfo {volume} {120}},\
  \bibinfo {pages} {183604} (\bibinfo {year} {2018})}\BibitemShut {NoStop}%
\bibitem [{\citenamefont {Hartmann}\ \emph
  {et~al.}(2020{\natexlab{b}})\citenamefont {Hartmann}, \citenamefont
  {Jenewein}, \citenamefont {Abend}, \citenamefont {Roura},\ and\ \citenamefont
  {Giese}}]{hartmann_atomic_2020}%
  \BibitemOpen
  \bibfield  {author} {\bibinfo {author} {\bibfnamefont {S.}~\bibnamefont
  {Hartmann}}, \bibinfo {author} {\bibfnamefont {J.}~\bibnamefont {Jenewein}},
  \bibinfo {author} {\bibfnamefont {S.}~\bibnamefont {Abend}}, \bibinfo
  {author} {\bibfnamefont {A.}~\bibnamefont {Roura}},\ and\ \bibinfo {author}
  {\bibfnamefont {E.}~\bibnamefont {Giese}},\ }\bibfield  {title} {\bibinfo
  {title} {Atomic raman scattering: Third-order diffraction in a double
  geometry},\ }\href {https://doi.org/10.1103/PhysRevA.102.063326} {\bibfield
  {journal} {\bibinfo  {journal} {Phys. Rev. A}\ }\textbf {\bibinfo {volume}
  {102}},\ \bibinfo {pages} {063326} (\bibinfo {year}
  {2020}{\natexlab{b}})}\BibitemShut {NoStop}%
\bibitem [{\citenamefont {Schubert}\ \emph {et~al.}(2021)\citenamefont
  {Schubert}, \citenamefont {Abend}, \citenamefont {Gersemann}, \citenamefont
  {Gebbe}, \citenamefont {Schlippert}, \citenamefont {Berg},\ and\
  \citenamefont {Rasel}}]{schubert_multi-loop_2021}%
  \BibitemOpen
  \bibfield  {author} {\bibinfo {author} {\bibfnamefont {C.}~\bibnamefont
  {Schubert}}, \bibinfo {author} {\bibfnamefont {S.}~\bibnamefont {Abend}},
  \bibinfo {author} {\bibfnamefont {M.}~\bibnamefont {Gersemann}}, \bibinfo
  {author} {\bibfnamefont {M.}~\bibnamefont {Gebbe}}, \bibinfo {author}
  {\bibfnamefont {D.}~\bibnamefont {Schlippert}}, \bibinfo {author}
  {\bibfnamefont {P.}~\bibnamefont {Berg}},\ and\ \bibinfo {author}
  {\bibfnamefont {E.~M.}\ \bibnamefont {Rasel}},\ }\bibfield  {title} {\bibinfo
  {title} {Multi-loop atomic {{Sagnac}} interferometry},\ }\href
  {https://doi.org/10.1038/s41598-021-95334-7} {\bibfield  {journal} {\bibinfo
  {journal} {Sci. Rep.}\ }\textbf {\bibinfo {volume} {11}},\ \bibinfo {pages}
  {16121} (\bibinfo {year} {2021})}\BibitemShut {NoStop}%
\bibitem [{\citenamefont {Sidorenkov}\ \emph {et~al.}(2020)\citenamefont
  {Sidorenkov}, \citenamefont {Gautier}, \citenamefont {Altorio}, \citenamefont
  {Geiger},\ and\ \citenamefont {Landragin}}]{sidorenkov_tailoring_2020}%
  \BibitemOpen
  \bibfield  {author} {\bibinfo {author} {\bibfnamefont {L.~A.}\ \bibnamefont
  {Sidorenkov}}, \bibinfo {author} {\bibfnamefont {R.}~\bibnamefont {Gautier}},
  \bibinfo {author} {\bibfnamefont {M.}~\bibnamefont {Altorio}}, \bibinfo
  {author} {\bibfnamefont {R.}~\bibnamefont {Geiger}},\ and\ \bibinfo {author}
  {\bibfnamefont {A.}~\bibnamefont {Landragin}},\ }\bibfield  {title} {\bibinfo
  {title} {{Tailoring Multiloop Atom Interferometers with Adjustable Momentum
  Transfer}},\ }\href {https://doi.org/10.1103/PhysRevLett.125.213201}
  {\bibfield  {journal} {\bibinfo  {journal} {Phys. Rev. Lett.}\ }\textbf
  {\bibinfo {volume} {125}},\ \bibinfo {pages} {213201} (\bibinfo {year}
  {2020})}\BibitemShut {NoStop}%
\bibitem [{\citenamefont {Mielec}\ \emph {et~al.}(2018)\citenamefont {Mielec},
  \citenamefont {Altorio}, \citenamefont {Sapam}, \citenamefont {Horville},
  \citenamefont {Holleville}, \citenamefont {Sidorenkov}, \citenamefont
  {Landragin},\ and\ \citenamefont {Geiger}}]{mielec_atom_2018}%
  \BibitemOpen
  \bibfield  {author} {\bibinfo {author} {\bibfnamefont {N.}~\bibnamefont
  {Mielec}}, \bibinfo {author} {\bibfnamefont {M.}~\bibnamefont {Altorio}},
  \bibinfo {author} {\bibfnamefont {R.}~\bibnamefont {Sapam}}, \bibinfo
  {author} {\bibfnamefont {D.}~\bibnamefont {Horville}}, \bibinfo {author}
  {\bibfnamefont {D.}~\bibnamefont {Holleville}}, \bibinfo {author}
  {\bibfnamefont {L.~A.}\ \bibnamefont {Sidorenkov}}, \bibinfo {author}
  {\bibfnamefont {A.}~\bibnamefont {Landragin}},\ and\ \bibinfo {author}
  {\bibfnamefont {R.}~\bibnamefont {Geiger}},\ }\bibfield  {title} {\bibinfo
  {title} {Atom interferometry with top-hat laser beams},\ }\href
  {https://doi.org/10.1063/1.5051663} {\bibfield  {journal} {\bibinfo
  {journal} {Appl. Phys. Lett.}\ }\textbf {\bibinfo {volume} {113}},\ \bibinfo
  {pages} {161108} (\bibinfo {year} {2018})}\BibitemShut {NoStop}%
\bibitem [{\citenamefont {Trimeche}\ \emph {et~al.}(2019)\citenamefont
  {Trimeche}, \citenamefont {Battelier}, \citenamefont {Becker}, \citenamefont
  {Bertoldi}, \citenamefont {Bouyer}, \citenamefont {Braxmaier}, \citenamefont
  {Charron}, \citenamefont {Corgier}, \citenamefont {Cornelius}, \citenamefont
  {Douch}, \citenamefont {Gaaloul}, \citenamefont {Herrmann}, \citenamefont
  {Müller}, \citenamefont {Rasel}, \citenamefont {Schubert}, \citenamefont
  {Wu},\ and\ \citenamefont {Pereira Dos~Santos}}]{trimeche_concept_2019}%
  \BibitemOpen
  \bibfield  {author} {\bibinfo {author} {\bibfnamefont {A.}~\bibnamefont
  {Trimeche}}, \bibinfo {author} {\bibfnamefont {B.}~\bibnamefont {Battelier}},
  \bibinfo {author} {\bibfnamefont {D.}~\bibnamefont {Becker}}, \bibinfo
  {author} {\bibfnamefont {A.}~\bibnamefont {Bertoldi}}, \bibinfo {author}
  {\bibfnamefont {P.}~\bibnamefont {Bouyer}}, \bibinfo {author} {\bibfnamefont
  {C.}~\bibnamefont {Braxmaier}}, \bibinfo {author} {\bibfnamefont
  {E.}~\bibnamefont {Charron}}, \bibinfo {author} {\bibfnamefont
  {R.}~\bibnamefont {Corgier}}, \bibinfo {author} {\bibfnamefont
  {M.}~\bibnamefont {Cornelius}}, \bibinfo {author} {\bibfnamefont
  {K.}~\bibnamefont {Douch}}, \bibinfo {author} {\bibfnamefont
  {N.}~\bibnamefont {Gaaloul}}, \bibinfo {author} {\bibfnamefont
  {S.}~\bibnamefont {Herrmann}}, \bibinfo {author} {\bibfnamefont
  {J.}~\bibnamefont {Müller}}, \bibinfo {author} {\bibfnamefont
  {E.}~\bibnamefont {Rasel}}, \bibinfo {author} {\bibfnamefont
  {C.}~\bibnamefont {Schubert}}, \bibinfo {author} {\bibfnamefont
  {H.}~\bibnamefont {Wu}},\ and\ \bibinfo {author} {\bibfnamefont
  {F.}~\bibnamefont {Pereira Dos~Santos}},\ }\bibfield  {title} {\bibinfo
  {title} {Concept study and preliminary design of a cold atom interferometer
  for space gravity gradiometry},\ }\href
  {https://doi.org/10.1088/1361-6382/ab4548} {\bibfield  {journal} {\bibinfo
  {journal} {Class. Quantum Grav.}\ }\textbf {\bibinfo {volume} {36}},\
  \bibinfo {pages} {215004} (\bibinfo {year} {2019})}\BibitemShut {NoStop}%
\bibitem [{\citenamefont {McDonald}\ \emph {et~al.}(2013)\citenamefont
  {McDonald}, \citenamefont {Kuhn}, \citenamefont {Bennetts}, \citenamefont
  {Debs}, \citenamefont {Hardman}, \citenamefont {Johnsson}, \citenamefont
  {Close},\ and\ \citenamefont {Robins}}]{mcdonald80bark2013}%
  \BibitemOpen
  \bibfield  {author} {\bibinfo {author} {\bibfnamefont {G.~D.}\ \bibnamefont
  {McDonald}}, \bibinfo {author} {\bibfnamefont {C.~C.~N.}\ \bibnamefont
  {Kuhn}}, \bibinfo {author} {\bibfnamefont {S.}~\bibnamefont {Bennetts}},
  \bibinfo {author} {\bibfnamefont {J.~E.}\ \bibnamefont {Debs}}, \bibinfo
  {author} {\bibfnamefont {K.~S.}\ \bibnamefont {Hardman}}, \bibinfo {author}
  {\bibfnamefont {M.}~\bibnamefont {Johnsson}}, \bibinfo {author}
  {\bibfnamefont {J.~D.}\ \bibnamefont {Close}},\ and\ \bibinfo {author}
  {\bibfnamefont {N.~P.}\ \bibnamefont {Robins}},\ }\bibfield  {title}
  {\bibinfo {title} {$80 \hbar k$ momentum separation with {{Bloch}}
  oscillations in an optically guided atom interferometer},\ }\href
  {https://doi.org/10.1103/PhysRevA.88.053620} {\bibfield  {journal} {\bibinfo
  {journal} {Phys. Rev. A}\ }\textbf {\bibinfo {volume} {88}},\ \bibinfo
  {pages} {053620} (\bibinfo {year} {2013})}\BibitemShut {NoStop}%
\bibitem [{\citenamefont {Schleich}(2001)}]{schleich_quantum_2001}%
  \BibitemOpen
  \bibfield  {author} {\bibinfo {author} {\bibfnamefont {W.~P.}\ \bibnamefont
  {Schleich}},\ }\href@noop {} {\emph {\bibinfo {title} {Quantum {O}ptics in
  {P}hase {S}pace}}}\ (\bibinfo  {publisher} {Wiley-VCH},\ \bibinfo {address}
  {Weinheim},\ \bibinfo {year} {2001})\BibitemShut {NoStop}%
\bibitem [{\citenamefont {Brion}\ \emph {et~al.}(2007)\citenamefont {Brion},
  \citenamefont {Pedersen},\ and\ \citenamefont
  {Mølmer}}]{brion_adiabatic_2007}%
  \BibitemOpen
  \bibfield  {author} {\bibinfo {author} {\bibfnamefont {E.}~\bibnamefont
  {Brion}}, \bibinfo {author} {\bibfnamefont {L.~H.}\ \bibnamefont
  {Pedersen}},\ and\ \bibinfo {author} {\bibfnamefont {K.}~\bibnamefont
  {Mølmer}},\ }\bibfield  {title} {\bibinfo {title} {Adiabatic elimination in
  a lambda system},\ }\href {https://doi.org/10.1088/1751-8113/40/5/011}
  {\bibfield  {journal} {\bibinfo  {journal} {J. Phys. A: Math. Theor.}\
  }\textbf {\bibinfo {volume} {40}},\ \bibinfo {pages} {1033} (\bibinfo {year}
  {2007})}\BibitemShut {NoStop}%
\bibitem [{\citenamefont {Bernhardt}\ and\ \citenamefont
  {Shore}(1981)}]{bernhardt_coherent_1981}%
  \BibitemOpen
  \bibfield  {author} {\bibinfo {author} {\bibfnamefont {A.~F.}\ \bibnamefont
  {Bernhardt}}\ and\ \bibinfo {author} {\bibfnamefont {B.~W.}\ \bibnamefont
  {Shore}},\ }\bibfield  {title} {\bibinfo {title} {Coherent atomic deflection
  by resonant standing waves},\ }\href
  {https://doi.org/10.1103/PhysRevA.23.1290} {\bibfield  {journal} {\bibinfo
  {journal} {Phys. Rev. A}\ }\textbf {\bibinfo {volume} {23}},\ \bibinfo
  {pages} {1290} (\bibinfo {year} {1981})}\BibitemShut {NoStop}%
\bibitem [{\citenamefont {Marte}\ and\ \citenamefont
  {Stenholm}(1992)}]{marte_multiphoton_1992}%
  \BibitemOpen
  \bibfield  {author} {\bibinfo {author} {\bibfnamefont {M.}~\bibnamefont
  {Marte}}\ and\ \bibinfo {author} {\bibfnamefont {S.}~\bibnamefont
  {Stenholm}},\ }\bibfield  {title} {\bibinfo {title} {Multiphoton resonances
  in atomic {Bragg} scattering},\ }\href {https://doi.org/10.1007/BF00325391}
  {\bibfield  {journal} {\bibinfo  {journal} {Appl. Phys. B: Lasers Opt.}\
  }\textbf {\bibinfo {volume} {54}},\ \bibinfo {pages} {443} (\bibinfo {year}
  {1992})}\BibitemShut {NoStop}%
\bibitem [{\citenamefont {MATLAB}(2021)}]{MATLAB:2021}%
  \BibitemOpen
  \bibfield  {author} {\bibinfo {author} {\bibnamefont {MATLAB}},\ }\href@noop
  {} {\emph {\bibinfo {title} {version 9.10.0 (R2021a)}}}\ (\bibinfo
  {publisher} {The MathWorks Inc.},\ \bibinfo {address} {Natick,
  Massachusetts},\ \bibinfo {year} {2021})\BibitemShut {NoStop}%
\bibitem [{\citenamefont {Shampine}\ and\ \citenamefont
  {Reichelt}(1997)}]{shampine_matlab_1997}%
  \BibitemOpen
  \bibfield  {author} {\bibinfo {author} {\bibfnamefont {L.~F.}\ \bibnamefont
  {Shampine}}\ and\ \bibinfo {author} {\bibfnamefont {M.~W.}\ \bibnamefont
  {Reichelt}},\ }\bibfield  {title} {\bibinfo {title} {The {MATLAB} {ODE}
  {Suite}},\ }\href {https://doi.org/10.1137/S1064827594276424} {\bibfield
  {journal} {\bibinfo  {journal} {SIAM J. Sci. Comput.}\ }\textbf {\bibinfo
  {volume} {18}},\ \bibinfo {pages} {1} (\bibinfo {year} {1997})}\BibitemShut
  {NoStop}%
\bibitem [{\citenamefont {Steck}(2001)}]{steck2001rubidium}%
  \BibitemOpen
  \bibfield  {author} {\bibinfo {author} {\bibfnamefont {D.~A.}\ \bibnamefont
  {Steck}},\ }\href {https://steck.us/alkalidata/} {\bibinfo {title} {Rubidium
  87 {D} line data}} (\bibinfo {year} {2001})\BibitemShut {NoStop}%
\bibitem [{\citenamefont {Perrin}\ \emph {et~al.}(2019)\citenamefont {Perrin},
  \citenamefont {Bernard}, \citenamefont {Bidel}, \citenamefont {Bonnin},
  \citenamefont {Zahzam}, \citenamefont {Blanchard}, \citenamefont {Bresson},\
  and\ \citenamefont {Cadoret}}]{Perrin2019}%
  \BibitemOpen
  \bibfield  {author} {\bibinfo {author} {\bibfnamefont {I.}~\bibnamefont
  {Perrin}}, \bibinfo {author} {\bibfnamefont {J.}~\bibnamefont {Bernard}},
  \bibinfo {author} {\bibfnamefont {Y.}~\bibnamefont {Bidel}}, \bibinfo
  {author} {\bibfnamefont {A.}~\bibnamefont {Bonnin}}, \bibinfo {author}
  {\bibfnamefont {N.}~\bibnamefont {Zahzam}}, \bibinfo {author} {\bibfnamefont
  {C.}~\bibnamefont {Blanchard}}, \bibinfo {author} {\bibfnamefont
  {A.}~\bibnamefont {Bresson}},\ and\ \bibinfo {author} {\bibfnamefont
  {M.}~\bibnamefont {Cadoret}},\ }\bibfield  {title} {\bibinfo {title}
  {Zero-velocity atom interferometry using a retroreflected frequency-chirped
  laser},\ }\href {https://doi.org/10.1103/PhysRevA.100.053618} {\bibfield
  {journal} {\bibinfo  {journal} {Phys. Rev. A}\ }\textbf {\bibinfo {volume}
  {100}},\ \bibinfo {pages} {053618} (\bibinfo {year} {2019})}\BibitemShut
  {NoStop}%
\bibitem [{\citenamefont {Bernard}\ \emph {et~al.}(2022)\citenamefont
  {Bernard}, \citenamefont {Bidel}, \citenamefont {Cadoret}, \citenamefont
  {Salducci}, \citenamefont {Zahzam}, \citenamefont {Schwartz}, \citenamefont
  {Bonnin}, \citenamefont {Blanchard},\ and\ \citenamefont
  {Bresson}}]{Bernard2022}%
  \BibitemOpen
  \bibfield  {author} {\bibinfo {author} {\bibfnamefont {J.}~\bibnamefont
  {Bernard}}, \bibinfo {author} {\bibfnamefont {Y.}~\bibnamefont {Bidel}},
  \bibinfo {author} {\bibfnamefont {M.}~\bibnamefont {Cadoret}}, \bibinfo
  {author} {\bibfnamefont {C.}~\bibnamefont {Salducci}}, \bibinfo {author}
  {\bibfnamefont {N.}~\bibnamefont {Zahzam}}, \bibinfo {author} {\bibfnamefont
  {S.}~\bibnamefont {Schwartz}}, \bibinfo {author} {\bibfnamefont
  {A.}~\bibnamefont {Bonnin}}, \bibinfo {author} {\bibfnamefont
  {C.}~\bibnamefont {Blanchard}},\ and\ \bibinfo {author} {\bibfnamefont
  {A.}~\bibnamefont {Bresson}},\ }\bibfield  {title} {\bibinfo {title} {Atom
  interferometry using $\sigma^+$-$\sigma^-$ {Raman} transitions between
  {$F=1,m_F=\mp1$} and {$F=2,m_F=\pm1$}},\ }\href
  {https://doi.org/10.1103/PhysRevA.105.033318} {\bibfield  {journal} {\bibinfo
   {journal} {Phys. Rev. A}\ }\textbf {\bibinfo {volume} {105}},\ \bibinfo
  {pages} {033318} (\bibinfo {year} {2022})}\BibitemShut {NoStop}%
\end{thebibliography}%

\end{document}